\documentclass[
    aps,
    pra,
    preprint,
    eqsecnum,
    amsmath,
    amssymb,
    nofootinbib
]{revtex4-2}

\usepackage{bm}
\usepackage{mathtools}
\usepackage{hyperref}
\usepackage{float}      
\usepackage{placeins} 
\usepackage{booktabs}
\usepackage{siunitx}

\usepackage{orcidlink}

\newcommand{\dd}{\mathrm{d}}
\newcommand{\ii}{\mathrm{i}}
\newcommand{\ee}{\mathrm{e}}
\newcommand{\eps}{\epsilon_0}

\begin{document}

\title{Quantum Optomechanics with Imperfect Mirrors and the Casimir--Polder Effect for Atoms with Different Dynamic Polarizabilities: A Unified Treatment via a Microscopic Model}

\author{Bergen Dahl\orcidlink{0009-0004-8171-7077}}
\email{bdahl@terpmail.umd.edu}
\affiliation{Department of Physics, University of Maryland, College Park, Maryland 20742, USA}
\author{Bei-Lok Hu \orcidlink{0000-0003-2489-9914}}
\email{blhu@umd.edu}
\affiliation{Joint Quantum Institute and Maryland Center for Fundamental Physics, University of Maryland, College Park, Maryland 20742, USA}
\date{August 21, 2026}

\begin{abstract}
This work brings the microscopic model of quantum optomechanics (QOM) proposed in \cite{Galley13} and developed in \cite{Sinha15,Lin18} one step closer to applicability under realistic experimental conditions, specifically those involving imperfect mirrors and real materials. The atom/mirror-oscillator-field (AMOF) model features an internal degree of freedom associated with a mirror or an atom. Its interaction with a quantum field determines either the transmission function of a mirror or the dynamic polarizability of an atom. We study three problems with this model: (1) an imperfect mirror moving in a cavity field, for which we compare the AMOF model with boundary-condition methods; (2) the ability of the AMOF dynamic polarizability to fit tabulated data at different frequencies; and (3) the quantum-fluctuation-induced Casimir--Polder energy between a dilute atomic half-space and a wall. We examine whether the three constituent parameters of the \textit{idf} oscillator can reproduce published results for a metastable He$^*$ atom near an Au plate and find excellent agreement. These examples show that the AMOF model has both a sound theoretical foundation, because it is based on the microscopic dynamics of its basic constituents, and considerable practical value, because it can produce accurate results for certain real materials.

\end{abstract}
\maketitle

\newpage
\tableofcontents
\newpage
\section{Introduction}

Fluctuation-induced quantum phenomena, though usually inconspicuously feeble, are of fundamental importance because they arise from ubiquitous quantum-field fluctuations. They appear across an enormous range of scales, from atomic structure---notably in van der Waals, Casimir, and Casimir--Polder forces \cite{LamorCasExp,MiltonAJP,IHACasPol,BuhmannI}---to galactic structure originating from quantum fluctuations amplified by the expansion of the universe \cite{GuthPi,HalHaw,CalHu95,Ambjorn}.

The Casimir force between two mirrors, or perfectly conducting plates, arises from the imbalance of vacuum energy density inside and outside the boundaries. The Casimir--Polder force between an atom and a conducting plate has the same origin in quantum-field fluctuations, except that one of the mirrors is replaced by an atom. These effects have been studied theoretically and experimentally for more than seven decades. Our aim is to bring a theoretical model proposed in 2013 for quantum optomechanics (QOM), suitable for studying the interaction of mirrors or atoms with a quantum field, closer to realistic experimental conditions involving imperfect or partially transmitting mirrors \cite{JaeReyImpMir} and real materials \cite{KMMCasReal} composed of atoms with different dynamic polarizabilities \cite{Clark}.

As quantum optomechanical experiments enter regimes in which measurement backaction, radiation-pressure noise, and microscopic material response become increasingly important, theoretical models must be refined to target specific observable phenomena. Standard optomechanical descriptions often encode the interaction between light and mechanical motion directly, for example by representing a mirror through imposed boundary conditions or by assuming the familiar $N\hat{x}$ coupling, where $N$ is the photon number and $\hat{x}$ is the mirror displacement. Such descriptions work well for large $N$ or sufficiently large systems in which dissipative heating can be mitigated. For smaller devices, however, dissipative heating can impair their operation. Conventional approaches may also become inadequate for questions of a quantum-information nature, such as entanglement between a mirror and a field \cite{Aspel,Genes,Sinha15}, for which photon-number fluctuations and correlations are important \cite{Caves}. These conventional semi-phenomenological models therefore require refinement to meet new challenges.

The Mirror-Oscillator-Field (MOF) model developed by Galley, Behunin, and Hu \cite{Galley13} takes such a microscopic approach by basing the description on a mirror's intrinsic dynamical properties rather than on externally imposed conditions. In addition to the mirror's mechanical degree of freedom (\textit{mdf}), which describes its center-of-mass motion, the MOF model introduces an internal degree of freedom (\textit{idf}) that couples to a quantum optical field, assumed for simplicity to be a scalar field $\Phi$. The \textit{idf}--field interaction determines the mirror's transmission and reflection functions. Depending on the parameter values, the model captures a range of optical responses, from broadband reflectivity to reflection within a narrow frequency band (see the plots in \cite{Sinha15} and the analysis in \cite{Lin18}).
More importantly, the three dynamical variables are determined simultaneously by a set of coupled equations, thereby ensuring self-consistency. This feature is important when treating measurement backaction and backreaction, including mirror cooling and the effect of dynamical-Casimir particle creation on the mirror's displacement. For further technical developments and applications of this model, see \cite{SLS21,Butera23,HHVacVis}.

The present work treats three problems with this model for a common purpose. The first concerns a slowly moving imperfect mirror in a cavity field, a system previously treated by Rego et al. using a Robin boundary condition imposed at the mirror's position \cite{Rego22}. Quantum optomechanics typically concerns a regime in which the mirror moves slowly enough that particle creation is negligible, or equivalently, in which particle number remains an adiabatic invariant. A fast-moving mirror that produces particles instead lies in the realm of the dynamical Casimir effect. The MOF model can treat both regimes; we will explore the latter in a subsequent paper. Here we treat this system using three intrinsic parameters: the coupling strength $\lambda$ between the \textit{idf} and the field, the \textit{idf} mass $m$ (the mass of a monopole interacting with a scalar field, or of the relevant electrons in the electromagnetic case), and the natural frequency $\Omega$ of the \textit{idf}, which is assumed to be harmonic.
We show how the reflection and transmission functions can be extracted from these parameters and matched to the results of Rego et al. The model is more versatile than a fixed boundary-condition description: by varying its parameters, we can move among perfectly reflecting, semitransparent, strong-coupling, and adiabatic regimes. This allows us to quantify scattering and particle production in terms of microscopic response properties.

We then turn our attention to atoms. When used to describe an atom, the mirror oscillator (mirosc) of the MOF model represents the atom's electronic response to an external field, which manifests as its dynamic polarizability. To connect the model with real atoms, we ask whether its three generic parameters are sufficient to reproduce the dynamic polarizabilities of different atomic species. We first motivate the connection between dynamic polarizability in simple electromagnetic models and in the MOF model, thereby deriving an effective dynamic polarizability in terms of the \textit{idf} parameters. We then determine how well this effective polarizability fits the frequency-dependent tabulated data of \cite{Derevianko_2010}. The agreement is very good. We therefore refer to this atomic extension of the microscopic model as the AMOF model, where ``A'' denotes atoms.

The third problem is the Casimir--Polder force between an atom and a mirror, analyzed using the AMOF model. This is a natural next step after modeling the atom with an appropriate dynamic polarizability and the imperfect mirror with suitable \textit{idf} dynamics. To apply the model to real materials, we use the results of Babb et al. \cite{Babb}. This application requires a model closer to electromagnetism: we employ a doublet of scalar fields to represent the two electromagnetic polarizations without introducing vector potentials and gauge constraints, together with a derivative coupling between the \textit{idf} and the field. We then compare our results with those of \cite{Babb} and find close agreement. Our interpretation is that the agreement arises because Ref. \cite{Babb} found that a single-oscillator model provides an excellent description of atoms in real materials, precisely the response supplied by the AMOF \textit{idf}.

The unifying theme of these three problems is that the AMOF model, based on the microscopic dynamics of three constituent variables, can effectively represent both the optical and kinematical properties of microscopic composite entities such as atoms and of macroscopic objects such as mirrors or dielectric materials (see, e.g., \cite{AFM2} and references therein). The successful treatment of the third problem provides evidence that the AMOF model is applicable to realistic materials.

This paper is organized as follows. In Sec.~II.A, we review the stationary-mirror setup described in detail in Ref.~\cite{Galley13}. In Sec.~II.B, we assign a trajectory $Z(t)$ to the mirror's center of mass and study particle production due to slow motion. We first express particle production in terms of the mirror's macroscopic optical properties and then rewrite the result in terms of dynamic polarizability to show how microscopic constituent properties affect the process. In Sec.~III, we demonstrate how different interactions in the AMOF model produce different optical responses. We compare the minimal $Q\Phi$ coupling with a bound charge coupled to an electric field, define an effective AMOF dynamic polarizability, and compare the single-oscillator model with the tabulated data of Ref.~\cite{Derevianko_2010}. In Secs.~III.D and III.E, we show how the derivative coupling $\dot Q\Phi$ reproduces the response of a thin sheet to an electric field.
Section IV uses the results of the previous two sections and a $(3+1)$-dimensional AMOF model to calculate the Casimir--Polder energy. Finally, Sec. V calculates corrections to the ideal Casimir--Polder free energy for a metastable He atom near an Au wall at $300~\mathrm{K}$ and compares them with the results of \cite{Babb} for real materials.

\section{Partial Reflectivity of an Imperfect Mirror from Its \textit{idf}'s Interaction with a Quantum Field}

We study an imperfect mirror in this section, first a stationary one to identify the three intrinsic constituent parameters in the MOF model, useful for matching the model to realistic mirrors and real atoms (next section).  We then study a moving mirror and show how the conventional method of imposing boundary conditions can be replaced by the dynamics of the mirror's internal degree of freedom (idf) and the quantum field. 

\subsection{Stationary Mirror: Three Intrinsic Constituent Parameters in the MOF Model}

We begin with the simplest case of a one-dimensional stationary mirror centered at $x=0$. This setup shows how two of the three dynamical variables in the MOF model, the \textit{idf} and the quantum field, enter the equations of motion. We thereby identify the three intrinsic constituent parameters central to applications of the MOF model involving imperfect mirrors and atoms with different dynamic polarizabilities, while also demonstrating the versatility of this microscopic model of QOM. For further discussion of an imperfect mirror in the MOF model and its relation to other common models of QOM, see \cite{Galley13} and the first half of \cite{Sinha15}.

The $(1+1)$-dimensional spacetime metric has the signature
\begin{equation}
    \eta_{\alpha\beta}
    =
    \mathrm{diag}(1,-1).
\end{equation}
We set $c=\hbar=1$ throughout this section.
The action describing the mirror-field interaction can be written as \cite{Galley13}
\begin{equation}\label{action}
    S[\Phi,Q]
    =
    \frac{1}{2}\int \dd^2x \,
    \partial_\alpha \Phi \, \partial^\alpha \Phi
    +
    \frac{m}{2}\int \dd t \,
    \left(\dot Q^2-\Omega^2Q^2\right)
    +
    \lambda\int \dd t \,
    Q(t)\Phi(t,0).
\end{equation}
Here $\Phi$ is a quantum scalar field, and $Q$ is the internal degree of freedom (\textit{idf}) of the imperfect mirror. The \textit{idf} is modeled as a harmonic oscillator (called a mirosc in \cite{Galley13}) with mass $m$ and natural frequency $\Omega$; for a two-level system, $\Omega$ would instead characterize the energy separation. The coupling constant $\lambda$ measures the interaction strength between the \textit{idf} and the scalar field. The complete MOF model also contains a mechanical degree of freedom (\textit{mdf}), namely the center-of-mass position $Z(t)$ of the mirror or atom. The field does not couple directly to the \textit{mdf}; rather, the \textit{mdf} affects the interaction because the \textit{idf} samples the field at the position of the mirror or atom. This dependence becomes important for moving mirrors or atoms, as in the dynamical Casimir and dynamical Casimir--Polder effects, and when the backaction of the field on the center-of-mass motion is of interest. Capturing the nonlinear interaction between the \textit{mdf} and \textit{idf} is one of the most challenging tasks in the microscopic dynamics. See, e.g., \cite{SLS21,HHVacVis} for case studies.

Varying with respect to $\Phi$ and $Q$ gives the following equations of motion
\begin{equation}\label{EOM1}
    \left(\partial_t^2-\partial_x^2\right)\Phi(t,x)
    =
    \lambda Q(t)\delta(x), 
\end{equation}
\begin{equation}
    m\ddot Q(t)+m\Omega^2Q(t)
    =
    \lambda\Phi(t,0).
\end{equation}
As a solution to Eq.~\eqref{EOM1}, consider a monochromatic wave incident from the left,
\begin{equation}\label{2.5}
    \Phi^L_\omega(t,x)
    =
    \ee^{-\ii\omega t}
    \left[
    \Theta(-x)
    \left(
    \ee^{\ii\omega x}
    +
    R(\omega)\ee^{-\ii\omega x}
    \right)
    +
    \Theta(x)T(\omega)\ee^{\ii\omega x}
    \right].
\end{equation}
We use the convention that $\Theta(0) = \frac{1}{2}$. 
The \textit{idf} is taken to respond at the same frequency,
\begin{equation}
    Q(t)
    =
    A\ee^{-\ii\omega t}.
\end{equation}
Its equation of motion then becomes
\begin{equation}
    m(\Omega^2-\omega^2)A
    =
    \lambda \Phi(t,0)\ee^{\ii\omega t}.
\end{equation}
Continuity of the field at $x=0$ requires
\begin{equation}
    \Phi(t,0^+)
    =
    \Phi(t,0^-).
\end{equation}
Using Eq.~\eqref{2.5}, this gives
\begin{equation}\label{2.9}
    T(\omega)
    =
    1+R(\omega).
\end{equation}
Therefore,
\begin{equation}\label{2.10}
    A
    =
    \frac{\lambda}{m(\Omega^2-\omega^2)}
    T(\omega).
\end{equation}
Integrating Eq.~\eqref{EOM1} across the origin yields the derivative boundary condition
\begin{equation}\label{2.11}
    \partial_x\Phi(t,0^+)-\partial_x\Phi(t,0^-)
    =
    \lambda Q(t).
\end{equation}
Using Eqs.~\eqref{2.9}--\eqref{2.11}, we find the reflection function
\begin{equation}
    R(\omega)
    =
    -
    \frac{\ii\lambda^2}
    {2m\omega(\Omega^2-\omega^2)+\ii\lambda^2}.
\end{equation}
The transmission function is
\begin{equation}
    T(\omega)
    =
    1+R(\omega)
    =
    \frac{2m\omega(\Omega^2-\omega^2)}
    {2m\omega(\Omega^2-\omega^2)+\ii\lambda^2}.
\end{equation}
The perfectly reflecting limit 
\begin{equation}
    R(\omega)\rightarrow -1,
    \qquad
    T(\omega)\rightarrow 0
\end{equation}
occurs in the strong \textit{idf}-field coupling regime ($\lambda \to \infty $), when the \textit{idf} is resonant with the field ($\Omega \to \omega$), or when the \textit{idf} mass is zero ($m \to 0$).   
The perfectly transmitting limit
\begin{equation}
    R(\omega)\rightarrow 0,
    \qquad
    T(\omega)\rightarrow 1 
\end{equation}
occurs in the weak \textit{idf}-field coupling regime ($\lambda \to 0$), when the \textit{idf} frequency is large ($\Omega \to \infty$), or when the \textit{idf} mass is large ($m \to \infty$). 

\subsection{Moving Mirror: Replacing Boundary Conditions with MOF Microdynamics}
We now illustrate how the MOF model can reproduce the function of a boundary condition imposed at a mirror's position. The model can adapt to different optical properties more readily than a fixed boundary condition. We introduce a mechanical degree of freedom (\textit{mdf}) $Z(t)$ describing the mirror's center-of-mass trajectory. We use the action in Eq.~\eqref{action}, with the \textit{idf}--field interaction now evaluated at $x=Z(t)$. The equations of motion become
\begin{equation}\label{EOM_at_Z}
    \left(\partial_t^2-\partial_x^2\right)\Phi(t,x)
    =
    \lambda Q(t)\delta(x - Z(t)), 
\end{equation}
\begin{equation}\label{mirosc_at_Z}
    m\ddot Q(t)+m\Omega^2Q(t)
    =
    \lambda\Phi(t,Z(t)).
\end{equation}

The physical effects of a moving imperfect mirror have been studied previously, notably in Ref.~\cite{Rego22}, which provides a useful comparison with the MOF treatment of the same problem.
Following Ref.~\cite{Rego22}, we split the scalar field into left and right regions:
\begin{equation}
    \Phi(t,x)
    =
    \Theta(Z(t)-x)\Phi_L(t,x)
    +
    \Theta(x-Z(t))\Phi_R(t,x).
\end{equation}
Continuity at the mirror interface requires 
\begin{equation}
    \Phi_L(t,Z(t))
    =
    \Phi_R(t,Z(t)).
\end{equation}
Integrating Eq.~\eqref{EOM_at_Z} across the mirror gives
\begin{equation}\label{derivative_bc}
    \dot Z(t)\Delta_Z(\partial_t\Phi)
    +
    \Delta_Z(\partial_x\Phi)
    =
    \lambda Q(t),
\end{equation}
where 
\begin{equation}
    \Delta_Z(\partial f) \equiv \partial f|_{Z - \epsilon} -  \partial f|_{Z + \epsilon}.
\end{equation}
Differentiating the continuity condition further constrains Eq.~\eqref{derivative_bc}, giving
\begin{equation}\label{2ndBC}
    \left(
    1-\dot Z(t)^2
    \right)
    \Delta_Z(\partial_x\Phi)
    =
    \lambda Q(t).
\end{equation}

\subsubsection{Incoming and Outgoing Modes}

We rewrite the field on the left and right of the mirror as a combination of ingoing and outgoing waves
\begin{equation}
    \Phi_L(t,x)
    =
    U_{\mathrm{in}}(t-x)
    +
    V_{\mathrm{out}}(t+x).
\end{equation}
\begin{equation}
    \Phi_R(t,x)
    =
    U_{\mathrm{out}}(t-x)
    +
    V_{\mathrm{in}}(t+x).
\end{equation}
Fourier transforming, 
\begin{equation}
    \Phi_L(t,x)
    =
    \int
    \frac{\dd\omega}{2\pi}
    \left[
    \widetilde U_{\mathrm{in}}(\omega)
    \ee^{-\ii\omega(t-x)}
    +
    \widetilde V_{\mathrm{out}}(\omega)
    \ee^{-\ii\omega(t+x)}
    \right].
\end{equation}
\begin{equation}
    \Phi_R(t,x)
    =
    \int
    \frac{\dd\omega}{2\pi}
    \left[
    \widetilde U_{\mathrm{out}}(\omega)
    \ee^{-\ii\omega(t-x)}
    +
    \widetilde V_{\mathrm{in}}(\omega)
    \ee^{-\ii\omega(t+x)}
    \right].
\end{equation}
Similarly, the oscillator is expanded as
\begin{equation}\label{4.10}
    Q(t)
    =
    \int
    \frac{\dd\omega}{2\pi}
    \widetilde Q(\omega)\ee^{-\ii\omega t}.
\end{equation}
Our continuity condition gives 
\begin{equation}
    \begin{aligned}
    &
    \int
    \frac{\dd\omega}{2\pi}
    \ee^{-\ii\omega t}
    \left[
    \widetilde U_{\mathrm{in}}(\omega)
    \ee^{\ii\omega Z(t)}
    +
    \widetilde V_{\mathrm{out}}(\omega)
    \ee^{-\ii\omega Z(t)}
    \right]
    \\
    &\hspace{2em}
    =
    \int
    \frac{\dd\omega}{2\pi}
    \ee^{-\ii\omega t}
    \left[
    \widetilde U_{\mathrm{out}}(\omega)
    \ee^{\ii\omega Z(t)}
    +
    \widetilde V_{\mathrm{in}}(\omega)
    \ee^{-\ii\omega Z(t)}
    \right].
    \end{aligned}
\end{equation}
We make the assumption that the displacement of $Z(t)$ is small
\begin{equation}
    \ee^{\pm \ii\omega Z(t)}
    =
    1\pm \ii\omega Z(t)+O(Z^2).
\end{equation}
For calculational ease, we define the following quantities
\begin{equation} \label{C(w)}
    C(\omega)
    =
    \widetilde U_{\mathrm{in}}(\omega)
    -
    \widetilde U_{\mathrm{out}}(\omega)
    +
    \widetilde V_{\mathrm{out}}(\omega)
    -
    \widetilde V_{\mathrm{in}}(\omega).
\end{equation}
\begin{equation}\label{D(w)}
    D(\omega)
    =
    \widetilde U_{\mathrm{out}}(\omega)
    -
    \widetilde U_{\mathrm{in}}(\omega)
    -
    \widetilde V_{\mathrm{in}}(\omega)
    +
    \widetilde V_{\mathrm{out}}(\omega).
\end{equation}
\begin{equation}\label{E(w)}
    E(\omega)
    =
    \widetilde U_{\mathrm{in}}(\omega)
    +
    \widetilde V_{\mathrm{out}}(\omega)
    +
    \widetilde U_{\mathrm{out}}(\omega)
    +
    \widetilde V_{\mathrm{in}}(\omega).
\end{equation}
\begin{equation}\label{F(w)}
    F(\omega)
    =
    \widetilde U_{\mathrm{in}}(\omega)
    -
    \widetilde V_{\mathrm{out}}(\omega)
    +
    \widetilde U_{\mathrm{out}}(\omega)
    -
    \widetilde V_{\mathrm{in}}(\omega).
\end{equation}
\begin{equation}
    \chi(\omega)
    =
    \frac{\lambda}
    {m(\Omega^2-\omega^2)}.
\end{equation}
The Fourier transform of the mirror trajectory is
\begin{equation}
    \widetilde Z(\omega-\omega')
    =
    \int
    \dd t \,
    \ee^{\ii(\omega-\omega')t}Z(t).
\end{equation}
To first order in $Z(t)$, continuity gives
\begin{equation}\label{C_cont}
    C(\omega)
    =
    \int
    \frac{\dd\omega'}{2\pi}
    \ii\omega'
    \widetilde Z(\omega-\omega')
    D(\omega').
\end{equation}

Using the second boundary condition, Eq.~\eqref{2ndBC}, dropping terms of order $Z^2$, and assuming $\dot Z^2\ll1$, we obtain
\begin{equation}\label{4.2}
    -\ii\omega D(\omega)
    =
    \lambda \widetilde Q(\omega).
\end{equation}
Using Eqs.~\eqref{4.2}, \eqref{4.10}, and \eqref{mirosc_at_Z}, we find
\begin{equation}\label{D=E}
    -
    \frac{\ii\omega}{\lambda}
    D(\omega)
    =
    \frac{\chi(\omega)}{2}
    \left[
    E(\omega)
    +
    \int
    \frac{\dd\omega'}{2\pi}
    \ii\omega'
    \widetilde Z(\omega-\omega')
    F(\omega')
    \right].
\end{equation}

\subsubsection{Zeroth-Order Scattering}

We expand the outgoing modes as a series of corrections in powers of $Z(t)$. 
\begin{equation}
    \widetilde U_{\mathrm{out}}(\omega)
    =
    \widetilde U_{\mathrm{out}}^{(0)}(\omega)
    +
    \widetilde U_{\mathrm{out}}^{(1)}(\omega)
    +
    \cdots,
\end{equation}
\begin{equation}
    \widetilde V_{\mathrm{out}}(\omega)
    =
    \widetilde V_{\mathrm{out}}^{(0)}(\omega)
    +
    \widetilde V_{\mathrm{out}}^{(1)}(\omega)
    +
    \cdots.
\end{equation}
We interpret $\widetilde U_{\mathrm{out}}^{(0)}(\omega)$ as the outgoing mode of the stationary mirror and $\widetilde U_{\mathrm{out}}^{(1)}(\omega)$ as the first-order correction due to mirror motion. Analogous interpretations hold for $\widetilde V_{\mathrm{out}}(\omega)$. We now order the equations in powers of $Z$, noting that the incoming data are of zeroth order.
At zeroth order, Eq. \ref{C(w)} becomes
\begin{equation}
    C^{(0)}(\omega)
    =
    0,
\end{equation}
which gives
\begin{equation}\label{4.27}
    \widetilde U_{\mathrm{in}}(\omega)
    -
    \widetilde U_{\mathrm{out}}^{(0)}(\omega)
    +
    \widetilde V_{\mathrm{out}}^{(0)}(\omega)
    -
    \widetilde V_{\mathrm{in}}(\omega)
    =
    0.
\end{equation}
Equation~\eqref{D=E} becomes
\begin{equation}\label{4.28}
    -
    \frac{\ii\omega}{\lambda}
    D^{(0)}(\omega)
    =
    \frac{\chi(\omega)}{2}
    E^{(0)}(\omega).
\end{equation}
Using the definitions of $D$ and $E$ together with Eqs.~\eqref{4.28} and \eqref{4.27},
\begin{equation}\label{scatter1}
    \widetilde U_{\mathrm{out}}^{(0)}(\omega)
    =
    T(\omega)\widetilde U_{\mathrm{in}}(\omega)
    +
    R(\omega)\widetilde V_{\mathrm{in}}(\omega),
\end{equation}
\begin{equation}\label{scatter2}
    \widetilde V_{\mathrm{out}}^{(0)}(\omega)
    =
    R(\omega)\widetilde U_{\mathrm{in}}(\omega)
    +
    T(\omega)\widetilde V_{\mathrm{in}}(\omega),
\end{equation}
with 
\begin{equation}
    T(\omega)
    =
    \frac{2\ii\omega}
    {2\ii\omega+\lambda\chi(\omega)}, \qquad R(\omega)
    =
    -
    \frac{\lambda\chi(\omega)}
    {2\ii\omega+\lambda\chi(\omega)}.
\end{equation}
Equivalently,
\begin{equation}\label{scatter_M}
    \begin{pmatrix}
    \widetilde U_{\mathrm{out}}^{(0)}(\omega)
    \\
    \widetilde V_{\mathrm{out}}^{(0)}(\omega)
    \end{pmatrix}
    =
    \begin{pmatrix}
    T(\omega) & R(\omega)
    \\
    R(\omega) & T(\omega)
    \end{pmatrix}
    \begin{pmatrix}
    \widetilde U_{\mathrm{in}}(\omega)
    \\
    \widetilde V_{\mathrm{in}}(\omega)
    \end{pmatrix}.
\end{equation}

\subsubsection{First-Order Correction}

At first order in $Z(t)$, Eq. \ref{C_cont} gives 
\begin{equation}
    C^{(1)}(\omega)
    =
    \int
    \frac{\dd\omega'}{2\pi}
    \ii\omega'
    \widetilde Z(\omega-\omega')
    D^{(0)}(\omega'),
\end{equation}
while Eq.~\eqref{C(w)} gives
\begin{equation}
    C^{(1)}(\omega)
    =
    \widetilde V_{\mathrm{out}}^{(1)}(\omega)
    -
    \widetilde U_{\mathrm{out}}^{(1)}(\omega).
\end{equation}
Using the above equations and Eq. \ref{scatter_M},
\begin{equation}\label{order 1}
    \begin{aligned}
    &
    \widetilde V_{\mathrm{out}}^{(1)}(\omega)
    -
    \widetilde U_{\mathrm{out}}^{(1)}(\omega)
    \\
    &\hspace{2em}
    =
    \int
    \frac{\dd\omega'}{2\pi}
    \ii\omega'
    \widetilde Z(\omega-\omega')
    2R(\omega')
    \left[
    \widetilde U_{\mathrm{in}}(\omega')
    +
    \widetilde V_{\mathrm{in}}(\omega')
    \right].
    \end{aligned}
\end{equation}
Using Eq.~\eqref{D=E}, the definitions of $F$, $D$, and $E$ to first order, and Eq.~\eqref{scatter_M}, we obtain
\begin{equation}\label{order2}
    \begin{aligned}
    &
    \widetilde V_{\mathrm{out}}^{(1)}(\omega)
    +
    \widetilde U_{\mathrm{out}}^{(1)}(\omega)
    \\
    &\hspace{2em}
    =
    2R(\omega)
    \int
    \frac{\dd\omega'}{2\pi}
    \ii\omega'
    \widetilde Z(\omega-\omega')
    \left[
    \widetilde U_{\mathrm{in}}(\omega')
    -
    \widetilde V_{\mathrm{in}}(\omega')
    \right].
    \end{aligned}
\end{equation}
Adding and subtracting Eqs.~\eqref{order 1} and \eqref{order2} gives
\begin{equation}
    \begin{aligned}
    \widetilde V_{\mathrm{out}}^{(1)}(\omega)
    &=
    \int
    \frac{\dd\omega'}{2\pi}
    \ii\omega'
    \widetilde Z(\omega-\omega')
    \\
    &\quad \times
    \left[
    \left(
    R(\omega)+R(\omega')
    \right)
    \widetilde U_{\mathrm{in}}(\omega')
    +
    \left(
    R(\omega')-R(\omega)
    \right)
    \widetilde V_{\mathrm{in}}(\omega')
    \right].
    \end{aligned}
\end{equation}
\begin{equation}
    \begin{aligned}
    \widetilde U_{\mathrm{out}}^{(1)}(\omega)
    &=
    \int
    \frac{\dd\omega'}{2\pi}
    \ii\omega'
    \widetilde Z(\omega-\omega')
    \\
    &\quad \times
    \left[
    \left(
    R(\omega)-R(\omega')
    \right)
    \widetilde U_{\mathrm{in}}(\omega')
    -
    \left(
    R(\omega)+R(\omega')
    \right)
    \widetilde V_{\mathrm{in}}(\omega')
    \right].
    \end{aligned}
\end{equation}

In matrix form,
\begin{equation}
    \begin{aligned}
    &
    \begin{pmatrix}
    \widetilde U_{\mathrm{out}}^{(1)}(\omega)
    \\
    \widetilde V_{\mathrm{out}}^{(1)}(\omega)
    \end{pmatrix}
    \\
    &=
    \int
    \frac{\dd\omega'}{2\pi}
    \ii\omega'
    \widetilde Z(\omega-\omega')
    \begin{pmatrix}
    R(\omega)-R(\omega') &
    -\left[
    R(\omega)+R(\omega')
    \right]
    \\
    R(\omega)+R(\omega') &
    R(\omega')-R(\omega)
    \end{pmatrix}
    \begin{pmatrix}
    \widetilde U_{\mathrm{in}}(\omega')
    \\
    \widetilde V_{\mathrm{in}}(\omega')
    \end{pmatrix}.
    \end{aligned}
\end{equation}

Thus the outgoing field may be written schematically as
\begin{equation}\label{Phi_out}
    \Phi_{\mathrm{out}}(\omega)
    =
    S_0(\omega)\Phi_{\mathrm{in}}(\omega)
    +
    \int
    \frac{\dd\omega'}{2\pi}
    S_1(\omega,\omega')
    \Phi_{\mathrm{in}}(\omega'),
\end{equation}
where 
\begin{equation}
    S_1(\omega,\omega') = i \omega' \tilde Z(\omega - \omega')  \begin{pmatrix}
    R(\omega)-R(\omega') &
    -\left[
    R(\omega)+R(\omega')
    \right]
    \\
    R(\omega)+R(\omega') &
    R(\omega')-R(\omega)
    \end{pmatrix}.
\end{equation}

\subsubsection{Particle Creation from a Slowly Moving Imperfect Mirror}

The time-dependent mirror trajectory mixes positive- and negative-frequency components. In quantum field theory, the backscattered component represents particle creation. We continue to use an $S$-matrix, or transition-amplitude, calculation and assume well-defined \textit{in} and \textit{out} states.
\begin{equation}
    \Phi_{\mathrm{out}}(\omega)
    =
    \int_0^\infty
    \frac{\dd\omega'}{2\pi}
    \left[
    S_1(\omega,\omega')
    \Phi_{\mathrm{in}}(\omega')
    +
    S_1(\omega,-\omega')
    \Phi_{\mathrm{in}}(-\omega')
    \right].
\end{equation}
Using the mode normalization
\begin{equation}
    \Phi_{\mathrm{in}}(\omega')
    =
    \frac{1}{\sqrt{2\omega'}}
    \hat a_{\omega'},\qquad \Phi_{\mathrm{in}}(-\omega')
    =
    \frac{1}{\sqrt{2\omega'}}
    \hat a^\dagger_{\omega'}
\end{equation}
we identify the Bogoliubov coefficient as
\begin{equation}
    \beta(\omega,\omega')
    =
    \sqrt{\frac{\omega}{\omega'}}
    S_1(\omega,-\omega'),
\end{equation}
whereby we obtain the particle number in the \textit{out} state as 
\begin{equation}
    N_{\mathrm{out}}(\omega)
    =
    \int_0^\infty
    \frac{\dd\omega'}{2\pi}
    \mathrm{Tr}
    \left[
    \beta^\dagger(\omega,\omega')
    \beta(\omega,\omega')
    \right].
\end{equation}
Using the first-order scattering kernel gives
\begin{align} \label{N_out}
    N_{\mathrm{out}}(\omega)
    &=
    4
    \int_0^\infty
    \frac{\dd\omega'}{2\pi}
    \omega\omega'
    \left|
    \widetilde Z(\omega+\omega')
    \right|^2
    \left[
    |R(\omega)|^2
    +
    |R(\omega')|^2
    \right]
\end{align}
From the above equation, we see that particle production occurs in pairs with frequencies $\omega$ and $\omega'$, while the reflection coefficients determine the efficiency with which the pairs couple to the internal oscillator. This efficiency is maximized for a perfectly reflecting mirror. An important advantage of the AMOF model for particle-production processes is that the optical response enters as a parameter-dependent function rather than as a boundary condition. This is the same response function obtained for the stationary mirror.

It is worth contrasting this result with other imperfect-mirror models, such as the $\delta-\delta'$ potential discussed in Refs.~\cite{Rego22,Rego16}. One advantage of that potential is its control over the phase of the reflection coefficient on either side of the mirror. The version of the AMOF model used here is parity symmetric, and its time dependence enters only through the trajectory. Consequently, it does not provide independent control of the left- and right-reflection phases. Such behavior could be incorporated microscopically by allowing $\lambda$ and $\Omega$ to differ on the two sides of the mirror. In its present form, however, the AMOF model is a microscopic dynamical model whose effective interface behavior is derived from internal degrees of freedom, whereas the $\delta-\delta'$ potential is an effective interface model by construction. The latter therefore provides no direct information about the mirror's material composition. The AMOF model also becomes transparent at arbitrarily high frequencies, as expected on physical grounds, whereas the $\delta-\delta'$ potential does not.

\subsubsection{Dynamical Casimir Effect with Tanh Trajectory}

For fast-moving mirrors, particle creation becomes important, and the system transitions from quantum optomechanics (QOM) to the dynamical Casimir effect (DCE). We use an example to show that particle production calculated with the MOF model agrees with results in the DCE literature. For simplicity, we consider a perfect mirror; the steps outlined above can also be used to obtain particle production by an imperfect AMOF mirror. A useful illustrative trajectory is \cite{BirDav82}
\begin{equation}
    Z(t)
    =
    A\tanh(\rho t),
\end{equation}
where $A$ is small. We also assume that $\dot Z(t)$ is small; for fixed $A$, this requires $\rho$ to be sufficiently small. For $\nu>0$,
\begin{equation}
    \widetilde Z(\nu)
    =
    \int_{-\infty}^{\infty}
    \dd t \,
    \ee^{\ii\nu t}
    A\tanh(\rho t).
\end{equation}
This gives
\begin{equation}
    \widetilde Z(\nu)
    =
    \frac{\ii\pi A}{\rho}
    \operatorname{csch}
    \left(
    \frac{\pi\nu}{2\rho}
    \right).
\end{equation}
For a perfectly reflecting mirror,
\begin{equation}
    |R(\omega)|^2
    =
    |R(\omega')|^2
    =
    1.
\end{equation}
Then
\begin{equation}
    N_{\mathrm{out}}(\omega)
    =
    \frac{8\pi^2A^2}{\rho^2}
    \int_0^\infty
    \frac{\dd\omega'}{2\pi}
    \omega\omega'
    \operatorname{csch}^2
    \left[
    \frac{\pi(\omega+\omega')}{2\rho}
    \right].
\end{equation}
Evaluating the integral gives
\begin{equation}
    N_{\mathrm{out}}(\omega)
    =
    \frac{16A^2\omega}{\pi}
    \ln
    \left[
    \frac{1}{1-\ee^{-\pi\omega/\rho}}
    \right].
\end{equation}
If we define the rescaled spectrum
\begin{equation}
    n_\omega
    =
    \frac{N_{\mathrm{out}}(\omega)}
    {A^2\rho},
\end{equation}
then
\begin{equation}
    n_\omega
    =
    \frac{16\omega}{\pi\rho}
    \ln
    \left[
    \frac{1}{1-\ee^{-\pi\omega/\rho}}
    \right].
\end{equation}
Defining 
\begin{equation}
    x = \frac{\pi\omega}{\rho},
\end{equation}
when the rise parameter $\rho$ is small, signifying a gradual expansion, $x\gg 1$,   we see that 
\begin{equation}
    n_\omega \sim \frac{16x}{\pi^2}\ee^{-x},
\end{equation}
so particle creation is exponentially suppressed.
\section{Atoms with Various Dynamic Polarizabilities in the AMOF Model}
The frequency-dependent electric polarizability of an atom is
\begin{equation}
    \alpha(u)
    =
    \sum_n'
    \frac{f_n}{\omega_n^2-u^2}
\end{equation}
where $f_n$ are the absorption oscillator strengths, $\omega_n$ are the excitation energies, and the primed sum includes both a sum over discrete states and an integral over continuum states \cite{Derevianko_2010,Dalgarno1966TheCO}. Analytically continuing to imaginary frequency, $u\rightarrow \ii\omega$,
gives
\begin{equation}\label{dp_analytic}
    \alpha(\ii\omega)
    =
    \sum_n'
    \frac{f_n}{\omega_n^2+\omega^2}.
\end{equation}
The tabulated dynamic polarizabilities in \cite{Derevianko_2010} combine empirical data with theoretical calculations based on Eq.~\eqref{dp_analytic}. To compare these data with the MOF model, we consider two couplings: the minimal $Q\Phi$ coupling and the derivative $\dot Q\Phi$ coupling. We seek a correspondence between the three intrinsic parameters of the MOF model introduced in Sec. II and the electric dynamic polarizabilities of different atoms. Because a model originally designed for an imperfect mirror can also describe an atom, we henceforth call its atomic extension the AMOF model and assign it an effective dynamic polarizability. This correspondence allows the AMOF model to describe the electric response of the atomic species of interest.

As a first example, Ref.~\cite{Galley13} established a correspondence between the MOF model and the Barton--Calogeracos (BC) model in the adiabatic regime, where $Q(t)$ evolves slowly:
\begin{equation}
    \left|\frac{\ddot Q}{\Omega^2Q}\right| \ll 1.
\end{equation}
It also established that the coupling constant $\lambda$ has the dimensions of charge density. Following Ref.~\cite{Sinha15}, we modify the action to
\begin{equation}
    S = \frac{\epsilon_0}{2c^2} \int d^2x (\partial_\alpha \Phi)^2 + \frac{m}{2} \int dt (\dot Q^2 - \Omega^2 Q^2) - \lambda \int dt Q(t) \Phi(t,0). 
\end{equation} 
In $(1+1)$ dimensions,
\begin{equation}
    [\epsilon_0] = \frac{\text{(charge)}^2 \text{(time)}^2}{\text{(mass)} \text{(length)}}, \qquad [\Phi] = \frac{\text{mass} \text{length}^2}{\text{time}^2 \text{charge}}, \qquad [Q] = \text{length}.
\end{equation}

\subsection{$Q\Phi$ Coupling}

For the minimal $Q\Phi$ coupling, the coupled \textit{idf}-field equations are 
\begin{equation}
    \frac{\eps}{c^2}
    \left(
    \partial_t^2\Phi
    -
    c^2\partial_x^2\Phi
    \right)
    =
    \lambda Q\delta(x),
\end{equation}
\begin{equation}
    m\ddot Q+m\Omega^2Q
    =
    \lambda\Phi(0,t).
\end{equation}
Using the ansatz in Eq.~\eqref{2.5}, the reflection function is
\begin{equation}
    R(\omega)
    =
    \frac{-\ii\lambda^2c}
    {\ii\lambda^2c+2m\omega\eps(\omega^2-\Omega^2)}.
\end{equation}
The reflectivity is
\begin{equation}
    |R|^2
    =
    \frac{\lambda^4c^2}
    {\lambda^4c^2+4m^2\omega^2\eps^2(\omega^2-\Omega^2)^2}.
\end{equation}
\subsection{Harmonic Oscillator Analog}
To connect the reflection function above with dynamic polarizability, we consider a simple example from electromagnetism. For a harmonically bound charge driven by an external electric field,
\begin{equation}
    m\ddot x+m\Omega_0^2x
    =
    qE(t), \qquad  E(t)
    =
    E_0\ee^{-\ii\omega t}.
\end{equation}
Taking
\begin{equation}
    x(t)
    =
    x(\omega)\ee^{-\ii\omega t},
\end{equation}
one obtains
\begin{equation}
    m(-\omega^2+\Omega_0^2)x(\omega)
    =
    qE_0.
\end{equation}
Thus,
\begin{equation}
    x(\omega)
    =
    \frac{q/m}{\Omega_0^2-\omega^2}
    E_0.
\end{equation}
The dipole moment is $p(\omega)= qx(\omega)$, and using $p(\omega)=\alpha(\omega)E_0$
gives
\begin{equation}
    \alpha(\omega)
    =
    \frac{q^2/m}{\Omega_0^2-\omega^2}.
\end{equation}
At imaginary frequency,
\begin{equation}
    \alpha(\ii\omega)
    =
    \frac{q^2/m}{\Omega_0^2+\omega^2}.
\end{equation}

\subsection{Effective Dynamic Polarizability in the AMOF Model}

From the $Q\Phi$ coupling, the reflectivity can be written as
\begin{equation}
    |R|^2
    =
    \frac{\lambda^4c^2}
    {\lambda^4c^2+4m^2\omega^2\eps^2(\omega^2-\Omega^2)^2}.
\end{equation}
Define the effective AMOF dynamic polarizability by
\begin{equation}
    \alpha_{\mathrm{MOF}}(\omega)
    =
    \frac{\lambda^2}
    {m(\Omega^2-\omega^2)}.
\end{equation}
The reflection function can then be rewritten in terms of dynamic polarizability as
\begin{equation}
    |R|^2
    =
    \frac{c^2\alpha_{\mathrm{MOF}}(\omega)^2}
    {c^2\alpha_{\mathrm{MOF}}(\omega)^2+4\eps^2\omega^2}.
\end{equation}
Under the analytic continuation $\omega\to\ii\xi$,
\begin{equation}
    \alpha_{\mathrm{MOF}}(\ii\xi)
    =
    \frac{\lambda^2}
    {m(\xi^2+\Omega^2)}.
\end{equation}
For Xe, He, Ar, Be, Li, and Cs, we fit the tabulated imaginary-frequency dynamic polarizabilities of Ref.~\cite{Derevianko_2010} to the form above. See Ref.~\cite{Dahl2026Code} for the relevant code.

We set the mass $m$ equal to the electron mass and treat the coupling $\lambda$ and internal oscillator frequency $\Omega$ as fit parameters. The parameters are obtained using a nonlinear least-squares procedure. In the results below, we employ logarithmic residuals,
\begin{equation}
    r_j=\log \alpha_{\mathrm{model}}(\ii\xi_j)-\log \alpha_{\mathrm{data}}(\ii\xi_j),
\end{equation}
rather than ordinary absolute residuals. This choice makes the fit sensitive to relative, or multiplicative, errors and prevents the large low-frequency polarizability values from dominating the optimization.
\begin{figure}[H]
    \centering
    \includegraphics[width=1.0\linewidth]{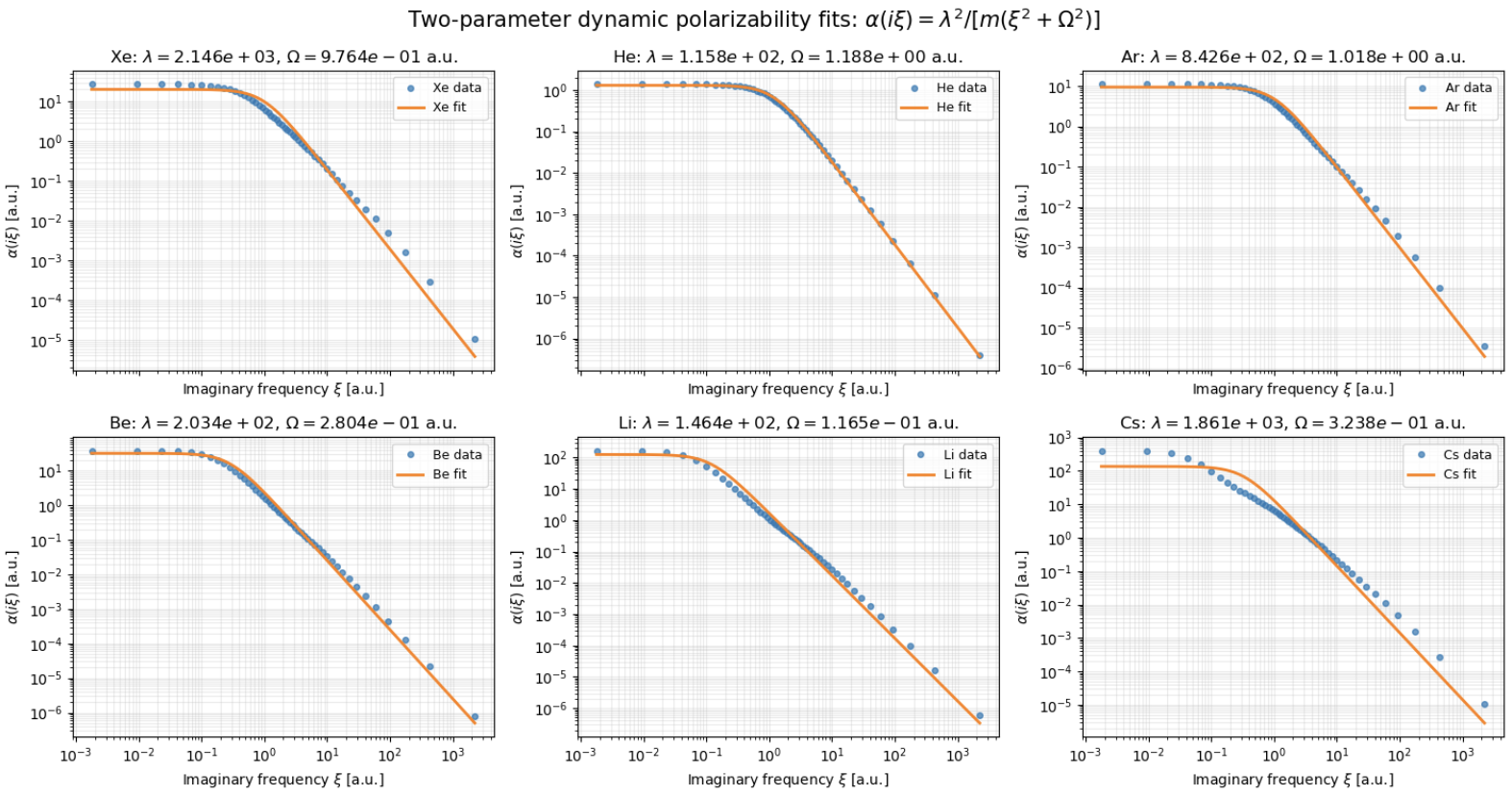}
    \caption{Fits of the dynamic polarizabilities for Xe, He, Ar, Be, Li, and Cs to the tabulated data.}
    \label{fig:polarizability-fits}
\end{figure}
\begin{figure}[H]
    \centering
    \includegraphics[width=1.0 \linewidth]{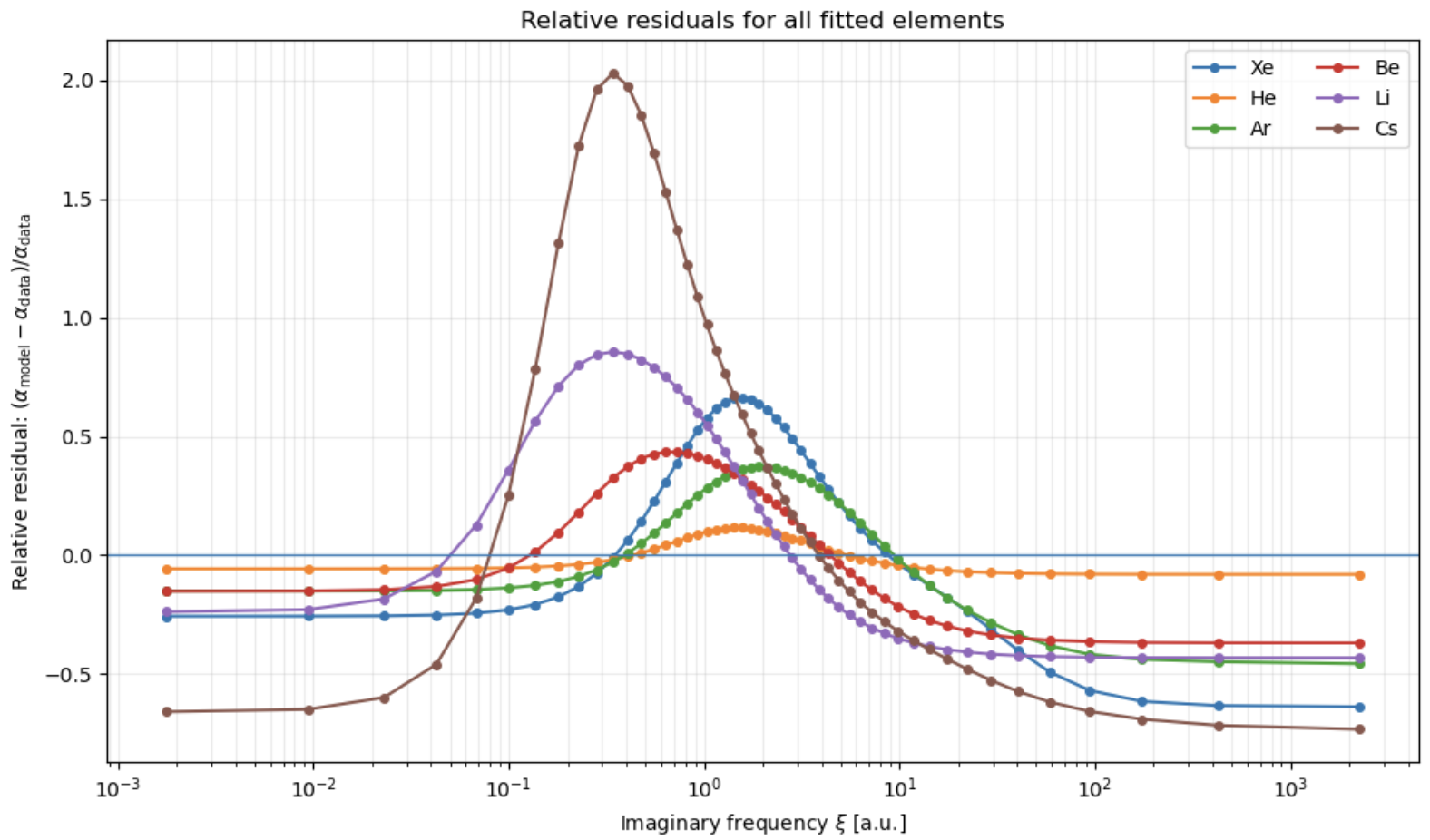}
    \caption{Relative residuals of the dynamic polarizabilities for Xe, He, Ar, Be, Li, and Cs.}
    \label{fig:relative-residuals}
\end{figure}
Several trends are visible in Fig.~\ref{fig:relative-residuals}. The AMOF model fits the lighter atoms more accurately, with the helium fit noticeably better than those of the other elements. We also find that the absence of unpaired valence electrons correlates with a better fit, as illustrated by the residuals for beryllium and argon. For heavy atoms such as cesium, the AMOF model reproduces the dynamic polarizability less accurately. As Eq.~\eqref{dp_analytic} indicates, a heavy atom's dynamic polarizability receives contributions from both lower- and higher-level electronic excitation channels. A response characterized by a single frequency $\Omega$ is therefore insufficient to capture the full behavior. Equation~\eqref{N_out} can also describe a moving atom with nonzero dynamic polarizability in the same manner as an imperfect moving mirror in the AMOF model. Given a trajectory, we can express particle production due to atomic motion entirely in terms of the dynamic polarizability.
\begin{equation}
    N_{\mathrm{out}}(\omega)
    =
    4
    \int_0^\infty
    \frac{\dd\omega'}{2\pi}
    \,
    \omega\omega'
    \left|
    \widetilde Z(\omega+\omega')
    \right|^2
    \left[
    \frac{\alpha_{\mathrm{MOF}}(\omega)^2}
    {4\omega^2+\alpha_{\mathrm{MOF}}(\omega)^2}
    +
    \frac{\alpha_{\mathrm{MOF}}(\omega')^2}
    {4\omega'^2+\alpha_{\mathrm{MOF}}(\omega')^2}
    \right].
\end{equation}

\subsection{$\dot Q\Phi$ Coupling}
We now explore the $\dot Q\Phi$ coupling. As in the previous section, the reflection function is (shown in \cite{Sinha15})
\begin{equation}
    R(\omega)
    =
    \frac{-\ii\lambda^2\omega}
    {\ii\lambda^2\omega+2m\eps c(\omega^2-\Omega^2)}, \qquad |R(\omega)|^2
    =
    \frac{\omega^2/\Omega^2}
    {\omega^2/\Omega^2+
    \left(
    2m\eps c\Omega/\lambda^2
    \right)^2
    \left(
    \omega^2/\Omega^2-1
    \right)^2}.
\end{equation}
Equivalently,
\begin{equation}
    |R(\omega)|^2
    =
    \frac{\eta^2}
    {\eta^2+r_p^2(\eta^2-1)^2}.
\end{equation}
With 
\begin{equation}
    \eta
    =
    \frac{\omega}{\Omega}, \qquad r_p
    =
    \frac{\Omega}{\Omega_P}
    =
    \frac{\Omega}{\lambda^2/(2m\eps c)},
\end{equation}
where we have defined the plasma frequency as
\begin{equation}
    \Omega_P
    =
    \frac{\lambda^2}{2m\eps c}.
\end{equation}
The three regimes that approximate a perfectly reflecting mirror are the same as in the $Q\Phi$ case:
\begin{equation}
    \lambda\rightarrow\infty,
    \qquad
    \omega=\Omega,
    \qquad
    m\rightarrow 0.
\end{equation}

\subsection{Thin Sheet Analog}

Whereas the $Q\Phi$ coupling models a single-oscillator response, we now show that the $\dot Q\Phi$ coupling can model a surface-current response. To motivate this result, we use a thin-sheet analog. For a thin mirror at $z=0$, take the induced surface polarization to be
\begin{equation}
    P_s(\omega)
    =
    n_s\alpha(\omega)E_{\parallel}(\omega).
\end{equation}
With the convention
\begin{equation}
    P_s(t)
    =
    P_s(\omega)\ee^{-\ii\omega t},
\end{equation}
the associated surface current is
\begin{equation}
    K(t)
    =
    \frac{\partial P_s(t)}{\partial t}.
\end{equation}
Therefore,
\begin{equation}
    K(\omega)
    =
    -\ii\omega P_s(\omega)
    =
    -\ii\omega n_s\alpha(\omega)E_{\parallel}(0).
\end{equation}
The magnetic field boundary condition is
\begin{equation}
    H_{\parallel}(0^+)-H_{\parallel}(0^-)
    =
    -K.
\end{equation}
For a normally incident plane wave in vacuum,
\begin{equation}
    H_{\mathrm{left}}
    =
    \frac{1}{Z_0}
    \left(
    E_i\ee^{\ii kz}
    -
    E_r\ee^{-\ii kz}
    \right), \qquad  H_{\mathrm{right}}
    =
    \frac{1}{Z_0}E_t\ee^{\ii kz}.
\end{equation}
Defining $Z_0=\sqrt{\mu_0/\eps}$, we write
\begin{equation}
    r
    =
    \frac{E_r}{E_i},
    \qquad
    t
    =
    \frac{E_t}{E_i},
\end{equation}
with $t=1+r$. Using the magnetic-field boundary condition,
\begin{equation}
    -\frac{2rE_i}{Z_0}
    =
    -\ii\omega n_s\alpha(\omega)E_i(1+r),
\end{equation}
and solving for the reflection amplitude gives
\begin{equation}
    r(\omega)
    =
    \frac{\ii Z_0\omega n_s\alpha(\omega)}
    {2-\ii Z_0\omega n_s\alpha(\omega)}.
\end{equation}
The physical reflectivity is
\begin{equation}
    |R|^2
    =
    \frac{\left[Z_0\omega n_s\alpha(\omega)\right]^2}
    {4+\left[Z_0\omega n_s\alpha(\omega)\right]^2}.
\end{equation}
Equivalently,
\begin{equation}
    |R|^2
    =
    \frac{1}
    {4/\left[Z_0\omega n_s\alpha(\omega)\right]^2+1}.
\end{equation}
Following the Barton--Calogeracos model, we interpret $\lambda$ as a charge density. This allows us to write
\begin{equation}
    \alpha(\omega)
    =
    \frac{q^2}{m(\Omega^2-\omega^2)} = \alpha_{\mathrm{MOF}}(\omega) N^2 
\end{equation}
where $N^2$ denotes the charge-carrier density in this parametrization. We define the corresponding plasma frequency as
\begin{equation}
    \Omega_P
    =
    \frac{\lambda^2}{2m\eps c}.
\end{equation}
We assume that $n_s\simeq N^2$, where $n_s$ is the surface charge-carrier density. Then, using $\lambda=qN$, one obtains
\begin{equation}
    \Omega_P
    =
    \frac{n_sq^2}{2m\eps c}.
\end{equation}
The reflectivity becomes
\begin{equation}
    |R|^2
    =
    \frac{1}
    {4c^2\eps^2m^2(\Omega^2-\omega^2)^2/(\omega n_sq^2)^2+1}.
\end{equation}
Equivalently,
\begin{equation}
    |R|^2
    =
    \frac{\Omega_P^2\omega^2}
    {\Omega^4(1-\omega^2/\Omega^2)^2+\Omega_P^2\omega^2}.
\end{equation}
Therefore,
\begin{equation}
    |R|^2
    =
    \frac{\eta^2}
    {\eta^2+r_p^2(1-\eta^2)^2}
\end{equation}
where $\eta$ and $r_p$ are defined as in the previous section. Thus, the material response described by the $\dot Q\Phi$ coupling is analogous to that of a macroscopic thin sheet.


\section{Casimir--Polder Effect Between an Atom and a Medium in the AMOF Model}
Our goal in the remainder of the paper is to apply the single-oscillator AMOF model to the Casimir--Polder effect. Using the parameters that govern the model, we seek to reproduce the theoretical calculations of Ref.~\cite{Babb} for the free energy of atoms near a plate under realistic conditions. The purpose of this calculation is not to propose an alternative to the Lifshitz formalism, but to demonstrate that the AMOF model reproduces the established atom–surface interaction starting directly from an action describing the field and the internal oscillator degrees of freedom. By integrating out these oscillators, we obtain an effective material response from which the corresponding reflection coefficients follow. Once expressed in terms of these coefficients, the resulting free energy naturally assumes the same scattering form used by Babb et al. In the dilute-atom limit, and within the single-oscillator approximation adopted here, agreement with their results is therefore expected rather than coincidental. The significance of this agreement lies not in reproducing the final Lifshitz calculation itself, but in showing that the chosen AMOF coupling generates the appropriate optical response and provides a consistent action-based route from the model’s microscopic oscillator parameters to the established macroscopic description of the atom–surface interaction.

We consider a half-space geometry in which the atomic medium occupies $-\infty<z<0$, the wall occupies $L<z<\infty$, and the intervening region is vacuum. We later take the atomic half-space to be dilute and isolate the contribution from a single atom. We label the atomic and wall regions by $A$ and $W$, respectively, and define
\begin{equation}
    \Theta_A(z)=\Theta(-z),
    \qquad
    \Theta_W(z)=\Theta(z-L).
\end{equation}
We take the internal oscillator fields to couple to the scalar field through the temporal-derivative interaction $\dot{Q}_a\Phi$, with $\lambda_A$ and $\lambda_W$ denoting the atom--field and wall--field coupling strengths, respectively. The Minkowski action is
\begin{align}
S={}&
\frac{\epsilon_0}{2}
\int dt\int d^3x\,
\left[
    (\partial_t\Phi)^2-c^2(\boldsymbol{\nabla}\Phi)^2
\right]
\nonumber\\
&+ \sum_{a = A,W}
n_a\int dt\int d^3x\,\Theta_a(z)
\left[
    \frac{m_a}{2}
    \left(
        \dot{Q}_a^2-\Omega_a^2Q_a^2
    \right)
    +\lambda_a\dot{Q}_a\Phi
\right]
\end{align}
where $\epsilon_0$ is the vacuum permittivity in $(3+1)$ dimensions, and $n_a$, with $a\in\{A,W\}$, is the number of \textit{idf} oscillators per unit volume in each material. To describe realistic bulk media, we promote the single \textit{idf} $Q$ to a continuum $Q(X)$ of independent local oscillators distributed throughout each half-space. Extending the \textit{idf} distribution across the transverse plane allows $\Phi$ to couple independently to $Q$ for every transverse-momentum mode. The resulting action leads to the Lifshitz formula.

Because the systems are stationary and in thermal equilibrium, we use finite-temperature quantum field theory and impose periodic boundary conditions in imaginary time $\tau$, with $\beta=1/(k_BT)$ and Euclidean-time period $\hbar\beta$.
Under a Wick rotation
\begin{equation}
    t\rightarrow-i\tau,
    \qquad
    \partial_t\rightarrow i\partial_\tau,
    \qquad
    S_E=-iS\big|_{t=-i\tau},
\end{equation}
the Euclidean action becomes
\begin{align}
S_E={}&
\frac{\epsilon_0}{2}
\int_0^{\hbar\beta}d\tau\int d^3x\,
\left[
    (\partial_\tau\Phi)^2
    +c^2(\boldsymbol{\nabla}\Phi)^2
\right]
\nonumber\\
&+ \sum_{a = A,W}
n_a\int_0^{\hbar\beta}d\tau\int d^3x\,\Theta_a(z)
\left[
    \frac{m_a}{2}
    \left(
        (\partial_\tau Q_a)^2+\Omega_a^2Q_a^2
    \right)
    -i\lambda_a(\partial_\tau Q_a)\Phi
\right]
\nonumber
\end{align}
The associated Euclidean equations of motion are
\begin{equation}
\epsilon_0
\left(
    -\partial_\tau^2-c^2\nabla^2
\right)
\Phi(\tau,\mathbf{x})
=
i\sum_{a\in\{A,W\}}
n_a\lambda_a
\partial_\tau Q_a(\tau,\mathbf{x})
\Theta_a(z),
\end{equation}
\begin{equation}
m_a
\left(
    -\partial_\tau^2+\Omega_a^2
\right)
Q_a(\tau,\mathbf{x})
=
-i\lambda_a\partial_\tau\Phi(\tau,\mathbf{x}).
\end{equation}
Fourier-transforming in $\tau$ and using the Matsubara frequencies
\begin{equation}
    \omega_n = \frac{2\pi n}{\hbar \beta}
\end{equation}
gives the effective equation of motion for $\Phi$,
\begin{equation}
    \left[
    \varepsilon(i\omega_n,z)\omega_n^2
    -c^2\nabla^2
\right]
\Phi_n(\mathbf{x})
=0
\end{equation}
with 
\begin{equation}
\varepsilon(i\omega_n,z)
=
1+
\chi_A(i\omega_n)\Theta(-z)
+
\chi_W(i\omega_n)\Theta(z-L), \qquad \chi_a(i\omega_n)
\equiv
\frac{n_a\alpha_a(i\omega_n)}{\epsilon_0},
\end{equation}
\begin{equation}
    \alpha_a(i\omega_n) = \frac{\lambda_a^2}{m_a (\omega_n^2 + \Omega_a^2)}.
\end{equation}
Here $\varepsilon$ is the effective material permittivity. The field $\Phi$ mimics the TE polarization. To reproduce both polarizations of the electromagnetic field, we introduce a second scalar field $\psi$ through the effective action term below (see the beginning of Ref.~\cite{Bhatt16}).
\begin{equation}
S_{\psi,\mathrm{int}}
=
\frac{1}{2\epsilon_0}
\sum_n\int d^3x\,
\left[
\frac{\omega_n^2}{c^2}\psi_{-n}\psi_n + 
\frac{1}{\varepsilon(i\omega_n,z)}\nabla\psi_n\cdot\nabla\psi_{-n}
\right]
.
\end{equation}
For a dilute atomic half-space, the TM polarization can be derived from an \textit{idf} description of $\psi$, analogous to that of $\Phi$, by adding an interaction proportional to $\psi\partial_iQ_\psi^i$ together with an appropriate free term. $Q_\psi$ represents a distinct \textit{idf} that couples to $\psi$ but has the same parameters as $Q$. This is to ensure that $\psi$ and $\Phi$ remain independent. Physically, this coupling represents a dipole interaction. Because the wall is not assumed to be dilute, deriving its exact TM reflection coefficient from the \textit{idf} interactions would additionally require a self-energy term proportional to $Q_\psi^2$.

We instead introduce the required TM scalar action directly because its physical motivation is clearer. The resulting equation of motion is
\begin{equation}
    \left[
    \frac{\omega_n^2}{c^2}
    -
    \nabla\cdot
    \frac{1}{\varepsilon(i\omega_n,z)}
    \nabla
    \right]\psi_n(\mathbf{x})=0.
\end{equation}
An ansatz for left-moving waves incident from the vacuum is
\begin{equation}
\Phi_n(z)
=
\begin{cases}
T_{TE, A} e^{\kappa_A z},
& z<0,
\\[4pt]
e^{\kappa_0 z}+R_{TE, A}e^{-\kappa_0 z},
& 0<z<L,
\end{cases}
\end{equation}
\begin{equation}
\psi_n(z)
=
\begin{cases}
T_{TM, A} e^{\kappa_A z},
& z<0,
\\[4pt]
e^{\kappa_0 z}+R_{TM, A}e^{-\kappa_0 z},
& 0<z<L,
\end{cases}
\end{equation}
where 
\begin{align}
\kappa_A(i\omega_n,k_\parallel)
&=
\sqrt{
k_\parallel^2+
\varepsilon_A(i\omega_n)\frac{\omega_n^2}{c^2}
},
\\
\kappa_0(i\omega_n,k_\parallel)
&=
\sqrt{
k_\parallel^2+\frac{\omega_n^2}{c^2}
},
\\
\kappa_W(i\omega_n,k_\parallel)
&=
\sqrt{
k_\parallel^2+
\varepsilon_W(i\omega_n)\frac{\omega_n^2}{c^2}
}.
\end{align}
Subscripts $A$ and $W$ identify the material region, while $\mathrm{TE}$ and $\mathrm{TM}$ identify the polarization. The boundary conditions are
\begin{equation}
    [\Phi_n] = 0, \qquad [\partial_z \Phi_n] = 0,
\end{equation}
\begin{align}
    [\psi_n] &= 0,\qquad \left[\frac{1}{\varepsilon}\partial_z\psi_n\right] = 0.
\end{align}
Using these boundary conditions and the equations of motion, we obtain the reflection coefficients
\begin{equation}\displaystyle
     R_{TE,a}
    =
    \frac{\kappa_0-\kappa_a}
    {\kappa_0+\kappa_a}, \qquad
    R_{TM, a}(i\omega_n,k_\parallel)
    =
    \frac{
    \varepsilon_a(i\omega_n)\kappa_0-\kappa_a
    }{\varepsilon_a(i\omega_n)\kappa_0+\kappa_a
    },
    \qquad a\in\{A,W\}.
\end{equation}
\section{Free Energy}
Starting from the action and using functional methods \cite{Ttira08}, we recover the Lifshitz free energy per unit area between the atomic half-space and the wall \cite{Babb}:
\begin{align}
\frac{F}{A}
=
k_B T
\sum_{n=0}^{\infty}{}'
\int\frac{d^2k_\parallel}{(2\pi)^2}
\Bigg[
\ln\left(
1-R_{TE, A} R_{TE, W} e^{-2\kappa_0L}
\right)
\nonumber+
\ln\left(
1-R_{TM, A} R_{TM, W} e^{-2\kappa_0L}
\right)
\Bigg].
\end{align}
In the dilute atomic-density limit, $n_A$ is small, and we therefore take
\begin{equation}
     \chi_A(i\omega_n)\ll1
\end{equation}
which gives the following approximations to the reflection coefficients:
\begin{equation}
    R_{TE, A}(i\omega_n,k_\parallel)
\simeq
-
\frac{n_A\lambda_A^2}
{4\epsilon_0m_A(\omega_n^2+\Omega_A^2)}
\frac{\omega_n^2/c^2}
{k_\parallel^2+\omega_n^2/c^2},
\end{equation}
\begin{equation}
    R_{TM, A}(i\omega_n,k_\parallel)
\simeq
\frac{n_A\lambda_A^2}
{4\epsilon_0m_A(\omega_n^2+\Omega_A^2)}
\frac{
2k_\parallel^2+\omega_n^2/c^2
}{
k_\parallel^2+\omega_n^2/c^2
}.
\end{equation}
An important feature is that the $n=0$ Matsubara mode does not vanish because of the TM contribution at the interface. This mode dominates the correction to the free energy at sufficiently large separations or high temperatures. Keeping only terms that are first order in $n_A$, we obtain
\begin{align}\label{totalFE}
\frac{F}{A}
=
-
\frac{n_A k_B T}{4\epsilon_0}
\sum_{n=0}^{\infty}{}'
\alpha_A(i\omega_n)
\int\frac{d^2k_\parallel}{(2\pi)^2}
\frac{e^{-2\kappa_0L}}{\kappa_0^2}
\times
\left[
\left(
2k_\parallel^2+\frac{\omega_n^2}{c^2}
\right) R_{TM, W}
-
\frac{\omega_n^2}{c^2} R_{TE, W}
\right].
\end{align}
The total free energy can be rewritten in terms of contributions from individual atoms:
\begin{equation}
\frac{F}{A}
=
n_A
\int_{-\infty}^{0}dz\,
F_{\mathrm{atom}}(L-z),
\end{equation}
which can be used to isolate the free energy of a single atom:
\begin{equation}
    F_{\mathrm{atom}}(L)
=
-\frac{1}{n_A}
\frac{\partial}{\partial L}
\left(\frac{F}{A}\right).
\end{equation}
Applying this formula to Eq.~\eqref{totalFE}, we find
\begin{equation}
F_{\mathrm{atom}}(L)
=
-
\frac{k_B T}{2\epsilon_0}
\sum_{n=0}^{\infty}{}'
\alpha_A(i\omega_n)
\int\frac{d^2k_\parallel}{(2\pi)^2}
\frac{e^{-2\kappa_0L}}{\kappa_0}
\times
\left[
\left(
2k_\parallel^2+\frac{\omega_n^2}{c^2}
\right)R_{TM, W}
-
\frac{\omega_n^2}{c^2}R_{TE, W}
\right].
\end{equation}
We obtain the zero-temperature Casimir--Polder energy for an atom near an ideal-metal wall by taking $\varepsilon\to\infty$, for which $R_{TM,W}=-R_{TE,W}=1$. At zero temperature,
\begin{equation}
    k_BT \sum_{n = 0}^\infty{}' \to \frac{\hbar}{2\pi} \int_0^\infty d\omega, 
\end{equation}
which gives the ideal Casimir--Polder energy of Ref.~\cite{Babb}, up to differences in unit conventions:
\begin{equation}
    E_\text{ideal} = -
\frac{\hbar}{4\pi\epsilon_0}
\int_0^\infty d\omega
\alpha_A(i\omega)
\int\frac{d^2k_\parallel}{(2\pi)^2}
\frac{e^{-2\kappa_0L}}{\kappa_0}
\times
\left[
\left(
2k_\parallel^2+\frac{\omega^2}{c^2}
\right)
+
\frac{\omega^2}{c^2}
\right].
\end{equation}
For $L\gg \frac{c}{\Omega_A}$, 
\begin{equation}
    E_\text{ideal} = - \frac{3\hbar c \alpha_A(0)}{32 \pi^2 \epsilon_0 L^4}.
\end{equation}

\subsection{Correction Factor to the Ideal Casimir--Polder Energy of a Metastable Helium Atom Near a Gold Wall}
\begin{figure}[ht]
    \centering
    \includegraphics[width=0.9\linewidth]{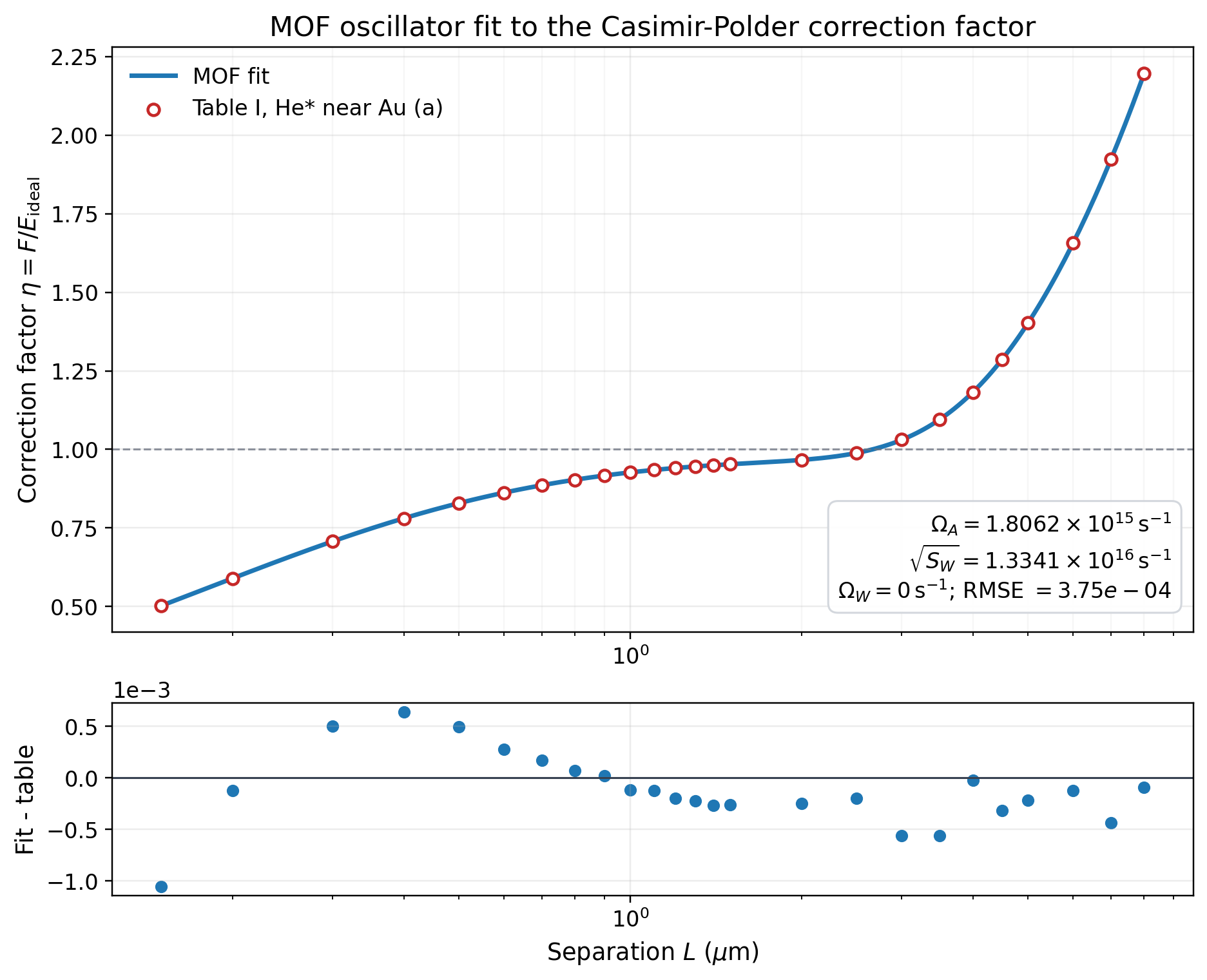}
    \caption{Fit of the correction factor as a function of distance to the data of Ref.~\cite{Babb}.}
    \label{fig:correction-factor}
\end{figure}

Our goal in this section is to use the single-atom free energy, expressed in terms of the AMOF parameters, to reproduce the calculations of Ref.~\cite{Babb} for helium near a gold wall at $300~\mathrm{K}$. Given
\begin{equation}
F_{\mathrm{atom}}(L)
=
-
\frac{k_B T}{2\epsilon_0}
\sum_{n=0}^{\infty}{}'
\alpha_A(i\omega_n)
\int\frac{d^2k_\parallel}{(2\pi)^2}
\frac{e^{-2\kappa_0L}}{\kappa_0}
\times
\left[
\left(
2k_\parallel^2+\frac{\omega_n^2}{c^2}
\right)R_{TM, W}
-
\frac{\omega_n^2}{c^2}R_{TE, W}
\right],
\end{equation}
we isolate the independent quantities that can be fitted to the data. Following Ref.~\cite{Babb}, we introduce the dimensionless variables
\begin{equation}
    y = 2L\kappa_0, \qquad \varpi_n = \frac{2L\omega_n}{c}.
\end{equation}
Then,
\begin{equation}
    F_\text{atom}(L) = - \frac{k_BT}{32\pi \epsilon_0 L^3} \sum_{n = 0}^\infty{}' \alpha_A(i\omega_n) \int_{\varpi_n}^\infty \dd y\,\ee^{-y} \left [
    (2y^2 - \varpi_n^2) R_{TM, W} - \varpi_n^2 R_{TE, W}
    \right ].
\end{equation}
We then have
\begin{equation}
    \eta = \frac{F}{E_\text{ideal}} = \frac{\pi k_B T L}{3 \hbar c} \sum_{n = 0}^\infty{}' \frac{\alpha_A(i\omega_n)}{\alpha_A(0)}\int_{\varpi_n}^\infty \dd y\,\ee^{-y} \left [
    (2y^2 - \varpi_n^2) R_{TM, W} - \varpi_n^2 R_{TE, W}
    \right ]. 
\end{equation}
where
\begin{equation}
    R_{TM,W} = \frac{\varepsilon_W(i\omega_n)y - \sqrt{y^2 + [\varepsilon_W(i\omega_n) - 1]\varpi_n^2}}{\varepsilon_W(i\omega_n)y + \sqrt{y^2 + [\varepsilon_W(i\omega_n) - 1]\varpi_n^2}},
\end{equation}
\begin{equation}
    R_{TE,W} = \frac{y - \sqrt{y^2 + [\varepsilon_W(i\omega_n) - 1]\varpi_n^2}}{y + \sqrt{y^2 + [\varepsilon_W(i\omega_n) - 1]\varpi_n^2}}.
\end{equation}
Performing the numerical integral and Matsubara sum, we fit the following three independent parameters:
\begin{equation}
    \Omega_A, \qquad \Omega_W, \qquad S_W = \frac{n_W\lambda_W^2}{\epsilon_0m_W}
\end{equation}
The resulting fit is shown in Fig.~\ref{fig:correction-factor}. The code can be found in Ref.~\cite{Dahl2026Code}. The fitted natural frequency of the helium atom is approximately $1.8\times10^{15}~\mathrm{rad\,s^{-1}}$. Reference~\cite{Babb} uses a single-oscillator frequency of $1.79\times10^{15}~\mathrm{rad\,s^{-1}}$, toward which the AMOF value converges. The quantity $\sqrt{S_W}$ can be interpreted as the wall's plasma frequency. The accepted plasma frequency of gold is approximately $\omega_p=1.37\times10^{16}~\mathrm{rad\,s^{-1}}$, and our fitted value lies within $3\%$ of it. This interpretation of $\sqrt{S_W}$ becomes clearer upon examining the wall permittivity:
\begin{equation}
    \varepsilon_W(i\omega_n) = 1 + \chi_W = 1 + \frac{n_W}{\epsilon_0} \frac{\lambda_W^2}{m_W(\omega_n^2 + \Omega_W^2)},
\end{equation}
The fit drives $\Omega_W\to0$, which yields
\begin{align}
    \varepsilon_W(i\omega_n) &= 1 + \frac{n_W}{\epsilon_0} \frac{\lambda_W^2}{m_W\omega_n^2} \\
    &= 1 +\frac{S_W}{\omega_n^2}
\end{align}
This is precisely the plasma model used in Ref.~\cite{Babb}.

\begin{table}[htbp]
    \centering
    \small
    \caption{Comparison of the correction factor from column (a) of Table~I in
    Ref.~\cite{Babb} with the fitted AMOF correction factor $\eta_{\mathrm{MOF}}$
    for metastable $\mathrm{He}^{*}$ near an Au wall at $T=300\,\mathrm{K}$.
    The residual is defined by
    $\Delta\eta=\eta_{\mathrm{MOF}}-\eta_{\text{Babb et al.}}$.}
    \label{tab:mof-paper-comparison}
    \renewcommand{\arraystretch}{1.05}
    \begin{tabular}{
        S[table-format=1.2]
        S[table-format=1.4]
        S[table-format=1.6]
        S[table-format=-1.6]
    }
        \toprule
        {$L\;(\si{\micro\meter})$}
        & {$\eta_{\text{Babb et al.}}$}
        & {$\eta_{\mathrm{MOF}}$}
        & {$\Delta\eta$} \\
        \midrule
        0.15 & 0.5039 & 0.502843 & -0.001057 \\
        0.20 & 0.5899 & 0.589776 & -0.000124 \\
        0.30 & 0.7070 & 0.707501 &  0.000501 \\
        0.40 & 0.7801 & 0.780740 &  0.000640 \\
        0.50 & 0.8285 & 0.828996 &  0.000496 \\
        0.60 & 0.8620 & 0.862275 &  0.000275 \\
        0.70 & 0.8859 & 0.886068 &  0.000168 \\
        0.80 & 0.9035 & 0.903566 &  0.000066 \\
        0.90 & 0.9167 & 0.916720 &  0.000020 \\
        1.00 & 0.9269 & 0.926779 & -0.000121 \\
        1.10 & 0.9347 & 0.934577 & -0.000123 \\
        1.20 & 0.9409 & 0.940698 & -0.000202 \\
        1.30 & 0.9458 & 0.945573 & -0.000227 \\
        1.40 & 0.9498 & 0.949530 & -0.000270 \\
        1.50 & 0.9531 & 0.952837 & -0.000263 \\
        2.00 & 0.9668 & 0.966553 & -0.000247 \\
        2.50 & 0.9889 & 0.988701 & -0.000199 \\
        3.00 & 1.0310 & 1.030434 & -0.000566 \\
        3.50 & 1.0960 & 1.095437 & -0.000563 \\
        4.00 & 1.1820 & 1.181972 & -0.000028 \\
        4.50 & 1.2860 & 1.285684 & -0.000316 \\
        5.00 & 1.4020 & 1.401783 & -0.000217 \\
        6.00 & 1.6560 & 1.655875 & -0.000125 \\
        7.00 & 1.9240 & 1.923559 & -0.000441 \\
        8.00 & 2.1960 & 2.195909 & -0.000091 \\
        \bottomrule
    \end{tabular}
\end{table}

\clearpage

\section{Conclusion}

The mirror-oscillator-field (MOF) model of quantum optomechanics (QOM), introduced in Ref.~\cite{Galley13}, describes the microscopic dynamics of a mirror's internal degree of freedom (\textit{idf}). This degree of freedom is modeled as a harmonic oscillator with mass $m$ and natural frequency $\Omega$ that interacts with an ambient quantum field through a coupling of strength $\lambda$. These three parameters characterize the microscopic constituents of systems such as imperfect mirrors \cite{Sinha15}, membranes, and individual atoms \cite{Lin18}, together with their coupling to an external field. We have shown that the same parameters can describe an atom's intrinsic electromagnetic response, which manifests as its dynamic polarizability. For lighter atoms, calculations using this atomic extension of the MOF model agree closely with the tabulated atomic data. This extension broadens the scope of the model to include atoms, motivating the name atom/mirror-oscillator-field (AMOF) model.

For a stationary mirror, we showed explicitly how $m$, $\Omega$, and $\lambda$ determine its frequency-dependent transparency and reflectivity. We then extended the analysis to a moving mirror, obtaining a microscopic and physically motivated model of particle creation by an imperfectly reflecting mirror. The resulting expression relates particle production to parameters that characterize the mirror's internal optical response. We compared this description with other models of imperfect reflection, such as the $\delta-\delta'$ potential. Rather than imposing a boundary condition, the AMOF model derives the effective interface behavior from the properties and dynamics of the mirror's microscopic constituents. Matching these parameters to a real material and solving the coupled equations for the \textit{mdf}, \textit{idf}, and field therefore provides a versatile description of realistic situations.

The ability of the \textit{idf} to describe atomic dynamics was developed further in Sec. III. Applying the AMOF model in $(1+1)$ dimensions, we expressed an atom's dynamic polarizability entirely in terms of the AMOF parameters and showed that a single-\textit{idf} oscillator can accurately reproduce the polarizabilities of lighter atoms. The minimal $Q\Phi$ interaction provides a direct scalar analog of a bound charge coupled to a field, while the derivative $\dot Q\Phi$ interaction demonstrates how an alternative microscopic coupling can reproduce the frequency-dependent response associated with an electric field. These results clarify how different microscopic interactions produce different effective optical properties in materials composed of such atoms.

Finally, in Secs. IV and V, we applied these scalar analogs to the Casimir--Polder interaction under conditions similar to those considered in Ref.~\cite{Babb}. We formulated the AMOF model in $(3+1)$ dimensions and represented the material bodies as half-spaces separated by vacuum. By taking one half-space to be a dilute atomic gas, we isolated the free energy of an individual atom interacting with the other surface. The model reproduced the correction to the ideal Casimir--Polder free energy found in Ref.~\cite{Babb} for a metastable helium atom near a gold wall at $T=300~\mathrm{K}$, with physically plausible fitted \textit{idf} parameters. This agreement is particularly encouraging because Ref.~\cite{Babb} found that a single-oscillator representation describes the exact Lifschitz atomic response to approximately $1\%$ accuracy.

Taken together, these results establish the AMOF model as a useful bridge between microscopic atomic or mirror-oscillator dynamics and macroscopic quantum-field-induced phenomena. The model describes frequency-dependent reflection, particle production by moving imperfect mirrors, atomic dynamic polarizability, and the finite-temperature Casimir--Polder interaction. It therefore provides a simple, reliable, and versatile framework connecting microscopic dynamics with the observable responses of realistic materials.

\begin{acknowledgments}
B.D. is supported by a Joint Quantum Institute summer undergraduate research fellowship. He acknowledges the use of OpenAI's ChatGPT 5.6 Sol for assistance with literature discovery, understanding Lifshitz theory, algebraic manipulations, coding, and technical presentation. All outputs were independently verified by the authors, who assume full responsibility for the content and accuracy of this work. We thank Prof. Shih-Yuin Lin for his interest in this project and for reading a preliminary draft of the moving-mirror calculation.
\end{acknowledgments}

\clearpage
\section{Appendix}
This appendix provides the algebraic steps required to derive the Lifshitz free energy from the AMOF action.
\subsection{Effective Action}
We first integrate out $Q$ and rewrite the theory as an effective action for $\Phi$. Starting from the Euclidean action, we integrate by parts, discard the boundary terms, and Fourier-transform in imaginary time:
\begin{align}
S_E={}&
\frac{1}{2}\sum_n \int d^3x \Phi_{-n} \epsilon_0( \omega_n^2 - c^2 \nabla^2) \Phi_n \\
&\qquad + \sum_{a,n} \int d^3x n_a \Theta_a(z) \left [ 
\frac{1}{2}m_a D_{a,n} Q_{a,-n} Q_{a,n} + \lambda_a \omega_n Q_{a,-n} \Phi_n 
\right ]
\end{align}
where $a\in\{A,W\}$ and
\begin{equation}
    D_{a,n} = \omega_n^2 + \Omega_a^2.
\end{equation}

Focusing on the $Q$-dependent part and simplifying the notation, we substitute the solution for $Q$ obtained above:
\begin{align}
    S_Q &= \sum_{n,a} n_a \int d^3x \Theta_a(z) \left[ -\frac{1}{2} \Phi_n \frac{\lambda_a^2 \omega_n^2}{m_a(\omega_n^2 + \Omega_a^2) } \Phi_{-n} + \frac{\lambda_a^2 \omega_n^2}{m_a (\omega_n^2 + \Omega_a^2) }  \Phi_n \Phi_{-n} 
   \right ] \\
    &= \sum_{n,a} n_a \int d^3x \Theta_a(z) \frac{1}{2} \Phi_n \frac{\lambda_a^2 \omega_n^2}{m_a(\omega_n^2 + \Omega_a^2) } \Phi_{-n}
\end{align}
Combining this result with the $\Phi$ part gives
\begin{equation}
    S_{\mathrm{eff}}[\Phi] = \frac{1}{2}\sum_n \int d^3x\,\Phi_{-n}(x) \mathbf M_n \Phi_n(x).
\end{equation}
Fourier transforming in transverse momenta gives 
\begin{equation}
    S_{\mathrm{eff}}[\Phi] = \frac{1}{2}\sum_n \int \frac{d^2k_\parallel}{(2\pi)^2}\int dz\,\Phi_{-n}(-k_\parallel,z) \mathbf M_n \Phi_n(k_\parallel,z).
\end{equation}
with 
\begin{equation}
    \mathbf M_n = \epsilon_0(\omega_n^2 + c^2 k_\parallel^2 - c^2\partial_z^2) + \sum_a n_a \Theta_a(z) \alpha_{a,\mathrm{MOF}}(\ii\omega_n)\omega_n^2.
\end{equation}
We define the susceptibility as 
\begin{equation}
    \chi_a(\ii\omega_n) = \frac{n_a}{\epsilon_0}\alpha_{a,\mathrm{MOF}}(\ii\omega_n)
\end{equation}
and 
\begin{equation}
\varepsilon(z,\ii\omega_n) = 1 + \sum_a \Theta_a(z) \chi_a(\ii\omega_n), \qquad \kappa_n^2 = k_\parallel^2 + \frac{\omega_n^2}{c^2}\varepsilon(z,\ii\omega_n).
\end{equation}
Then, 
\begin{equation}
    \mathbf M_n = \epsilon_0c^2 \left[
-\partial_z^2 + \kappa_n^2(z)
    \right].
\end{equation}

The free energy takes the form
\begin{equation}
    F = -k_B T \ln(Z). 
\end{equation}
The normalized partition function is
\begin{equation}
    Z = \prod_{n,\mathbf{k}_\parallel} \left[\frac{\det(\mathbf M_n)}{\det(\mathbf M_{0,n})}\right]^{-1/2}
\end{equation}
where $\mathbf M_{0,n}$ is the free operator for Matsubara mode $n$. Thus, the normalized free energy per unit area is
\begin{equation}
    \mathcal F = \frac{k_BT}{2}\int \frac{d^2k_\parallel}{(2\pi)^2}\sum_n \ln\left(\frac{\det(\mathbf M_n)}{\det(\mathbf M_{0,n})}\right).
\end{equation}
This expression contains ultraviolet divergences. We renormalize it by defining a reference configuration in which the atom and wall are at infinite separation and subtracting that contribution at the end of the calculation.

\subsection{Functional Determinant}
To evaluate the determinant, we follow Ref.~\cite{Ttira11} and introduce an auxiliary function $\psi$ to apply the Gel'fand--Yaglom theorem. The key relation is
\begin{equation}
    \frac{\det(\mathbf M_n)}{\det(\mathbf M_{0,n})} = \frac{\psi_W(z_+)}{\psi_{\mathrm{free}}(z_+)}.
\end{equation}
The task is to determine $\psi$. We first place the system in a box with
\begin{equation}
    z_- = -d_A, \qquad z_+ = L + d_W.
\end{equation}
$\psi $ is defined by 
\begin{equation}
    [- \partial_z^2 + \kappa_n^2 ] \psi(z) = 0. 
\end{equation}
This yields three ordinary differential equations:
\begin{equation}
    \psi_A^{''} = \kappa_A^2 \psi_A, \qquad \psi_0^{''} = \kappa_0^2 \psi_0, \qquad \psi_W^{''} = \kappa_W^2 \psi_W.
\end{equation}
We impose the initial conditions
\begin{equation}
    \psi(z_-) = 0, \qquad \psi'(z_-) = 1. 
\end{equation}
In the dilute atomic-density region, 
\begin{equation}
    \psi_A'' - \kappa_A^2 \psi_A = 0.
\end{equation}
The general solution is
\begin{equation}
    \psi_A(z) = C_A \sinh[\kappa_A(z + d_A)] + D_A \cosh[\kappa_A(z + d_A)].
\end{equation}
Applying the boundary conditions gives
\begin{equation}
    \psi_A(z) = \frac{1}{\kappa_A} \sinh[\kappa_A(z + d_A)], \qquad \psi'_A(z) = \cosh[\kappa_A(z+d_A)]
\end{equation}
The initial data for the vacuum solution are therefore
\begin{equation}
    \psi_A(0) = \frac{\sinh(\kappa_Ad_A)}{\kappa_A}, \qquad \psi_A'(0) = \cosh(\kappa_Ad_A).
\end{equation}
To find $\psi(z_+)$, we propagate the initial data toward increasing $z$. The solution at any point $z_f>z_i$ is uniquely determined by
\begin{equation}
    Y(z_f) = P_i(z_f - z_i) Y(z_i)
\end{equation}
with 
\begin{equation} 
   P_i(\Delta z) = \begin{pmatrix}\displaystyle
       \cosh(\kappa_i \Delta z) &\displaystyle \frac{1}{\kappa_i} \sinh(\kappa_i \Delta z) \\\displaystyle
       \kappa_i \sinh(\kappa_i \Delta z) & \cosh(\kappa_i \Delta z)\displaystyle
   \end{pmatrix}
   ,\qquad Y(z) = \begin{pmatrix}
       \displaystyle \psi(z) \\
        \psi'(z) \displaystyle
    \end{pmatrix}. 
\end{equation}
Then, for $0<z<L$, the vacuum solution is $Y(z) = P_0(z) Y(0)$. 
Thus, 
\begin{align}
    \psi_0(z ) \displaystyle&= \displaystyle\cosh(\kappa_0z) \psi_A(0) + \frac{1}{\kappa_0} \sinh(\kappa_0z) \psi_A'(0) \\\displaystyle
    &= \cosh(\kappa_0z) \frac{1}{\kappa_A} \sinh[\kappa_A(d_A)] + \frac{1}{\kappa_0} \sinh(\kappa_0z) \cosh[\kappa_Ad_A]
\end{align}
At the wall interface $z = L$, 
\begin{align}
\psi_0(L)
&=
\frac{\sinh(\kappa_A d_A)}{\kappa_A}
\cosh(\kappa_0 L)
+
\frac{\cosh(\kappa_A d_A)}{\kappa_0}
\sinh(\kappa_0 L),
\\[1ex]
\psi_0'(L)
&=
\frac{\kappa_0}{\kappa_A}
\sinh(\kappa_A d_A)\sinh(\kappa_0 L)
+
\cosh(\kappa_A d_A)\cosh(\kappa_0 L).
\end{align}

We propagate once more to find $\psi$ in the wall region,
$L<z<L+d_W$. Thus,
\begin{align}
\psi_W(z)
&=
\cosh\!\bigl[\kappa_W(z-L)\bigr]\psi_0(L)
+
\frac{1}{\kappa_W}
\sinh\!\bigl[\kappa_W(z-L)\bigr]\psi_0'(L),
\\[1ex]
\psi_W'(z)
&=
\kappa_W
\sinh\!\bigl[\kappa_W(z-L)\bigr]\psi_0(L)
+
\cosh\!\bigl[\kappa_W(z-L)\bigr]\psi_0'(L).
\end{align}
The full solution in the wall is
\begin{align}
\psi_W(z)
={}&
\cosh\!\bigl[\kappa_W(z-L)\bigr]
\left[
\frac{\sinh(\kappa_A d_A)}{\kappa_A}
\cosh(\kappa_0 L)
+
\frac{\cosh(\kappa_A d_A)}{\kappa_0}
\sinh(\kappa_0 L)
\right]
\nonumber\\
&+
\frac{\sinh\!\bigl[\kappa_W(z-L)\bigr]}{\kappa_W}
\left[
\frac{\kappa_0}{\kappa_A}
\sinh(\kappa_A d_A)\sinh(\kappa_0 L)
+
\cosh(\kappa_A d_A)\cosh(\kappa_0 L)
\right].
\end{align}
For the noninteracting solution,
\begin{equation}
    \psi_{\mathrm{free}}(z_+)
    =
   \frac{1}{\kappa_0}
    \sinh\!\bigl[\kappa_0(d_A+L+d_W)\bigr]
\end{equation}
We now evaluate
\begin{equation}
    \frac{\psi_W(z_+)}{\psi_{\mathrm{free}}(z_+)}
\end{equation}
for $d_W, d_A \to \infty$.

At the right boundary,
\begin{equation}
    z_+=L+d_W,
    \qquad
    z_+-L=d_W.
\end{equation}
Therefore,
\begin{align}
\psi_W(z_+)
={}&
\cosh(\kappa_W d_W)
\left[
\frac{\sinh(\kappa_A d_A)}{\kappa_A}
\cosh(\kappa_0L)
+
\frac{\cosh(\kappa_A d_A)}{\kappa_0}
\sinh(\kappa_0L)
\right]
\nonumber\\
&+
\frac{\sinh(\kappa_Wd_W)}{\kappa_W}
\left[
\frac{\kappa_0}{\kappa_A}
\sinh(\kappa_Ad_A)\sinh(\kappa_0L)
+
\cosh(\kappa_Ad_A)\cosh(\kappa_0L)
\right].
\end{align}

In the half-space limit $d_A,d_W\to\infty$, we use
\begin{equation}
    \sinh(\kappa_A d_A)
    \sim
    \cosh(\kappa_A d_A)
    \sim
    \frac{1}{2}e^{\kappa_A d_A},
\end{equation}
and
\begin{equation}
    \sinh(\kappa_W d_W)
    \sim
    \cosh(\kappa_W d_W)
    \sim
    \frac{1}{2}e^{\kappa_W d_W}.
\end{equation}
It follows that
\begin{align}
\psi_W(z_+)
\sim{}&
\frac{e^{\kappa_A d_A+\kappa_Wd_W}}{4}
\left[
\left(
\frac{1}{\kappa_A}
+
\frac{1}{\kappa_W}
\right)
\cosh(\kappa_0L)
\right.
\nonumber\\
&\hspace{37mm}\left.
+
\left(
\frac{1}{\kappa_0}
+
\frac{\kappa_0}{\kappa_A\kappa_W}
\right)
\sinh(\kappa_0L)
\right].
\end{align}

Using
\begin{equation}
    \cosh(\kappa_0L)
    =
    \frac{e^{\kappa_0L}+e^{-\kappa_0L}}{2},
    \qquad
    \sinh(\kappa_0L)
    =
    \frac{e^{\kappa_0L}-e^{-\kappa_0L}}{2},
\end{equation}
we obtain
\begin{align}
\psi_W(z_+)
\sim{}&
\frac{e^{\kappa_A d_A+\kappa_Wd_W}}
{8\kappa_A\kappa_0\kappa_W}
\Big[
(\kappa_A+\kappa_0)(\kappa_W+\kappa_0)
e^{\kappa_0L}
\nonumber\\
&\hspace{42mm}
-
(\kappa_A-\kappa_0)(\kappa_W-\kappa_0)
e^{-\kappa_0L}
\Big].
\end{align}
Equivalently,
\begin{align}
\psi_W(z_+)
\sim{}&
\frac{(\kappa_A+\kappa_0)(\kappa_W+\kappa_0)}
{8\kappa_A\kappa_0\kappa_W}
e^{\kappa_A d_A+\kappa_Wd_W+\kappa_0L}
\nonumber\\
&\times
\left[
1-
\frac{(\kappa_A-\kappa_0)(\kappa_W-\kappa_0)}
{(\kappa_A+\kappa_0)(\kappa_W+\kappa_0)}
e^{-2\kappa_0L}
\right].
\end{align}
The noninteracting reference solution is
\begin{equation}
    \psi_{\mathrm{free}}(z_+)
    =
    \frac{1}{\kappa_0}
    \sinh\!\left[
    \kappa_0(d_A+L+d_W)
    \right].
\end{equation}
As $d_A,d_W\to\infty$,
\begin{equation}
    \psi_{\mathrm{free}}(z_+)
    \sim
    \frac{1}{2\kappa_0}
    e^{\kappa_0(d_A+L+d_W)}.
\end{equation}
Therefore,
\begin{align}
\frac{\psi_W(z_+)}
{\psi_{\mathrm{free}}(z_+)}
\sim{}&
\frac{(\kappa_A+\kappa_0)(\kappa_W+\kappa_0)}
{4\kappa_A\kappa_W}
e^{(\kappa_A-\kappa_0)d_A}
e^{(\kappa_W-\kappa_0)d_W}
\nonumber\\
&\times
\left[
1-
\frac{(\kappa_A-\kappa_0)(\kappa_W-\kappa_0)}
{(\kappa_A+\kappa_0)(\kappa_W+\kappa_0)}
e^{-2\kappa_0L}
\right].
\end{align}
The reflection coefficients are
\begin{equation}
    r_A
    =
    \frac{\kappa_0-\kappa_A}
    {\kappa_0+\kappa_A},
    \qquad
    r_W
    =
    \frac{\kappa_0-\kappa_W}
    {\kappa_0+\kappa_W}.
\end{equation}
Their product is
\begin{equation}
    r_A r_W
    =
    \frac{(\kappa_A-\kappa_0)(\kappa_W-\kappa_0)}
    {(\kappa_A+\kappa_0)(\kappa_W+\kappa_0)}.
\end{equation}
Hence,
\begin{align}
\frac{\psi_W(z_+)}
{\psi_{\mathrm{free}}(z_+)}
\sim{}&
\mathcal{C}(d_A,d_W)
\left[
1-r_A r_W e^{-2\kappa_0L}
\right],
\end{align}
where
\begin{equation}
    \mathcal{C}(d_A,d_W)
    =
    \frac{(\kappa_A+\kappa_0)(\kappa_W+\kappa_0)}
    {4\kappa_A\kappa_W}
    e^{(\kappa_A-\kappa_0)d_A}
    e^{(\kappa_W-\kappa_0)d_W}.
\end{equation}
At infinite separation,
\begin{equation}
    \frac{\psi_W(z_+;\infty)}{\psi_{\mathrm{free}}(z_+)}
    =
    \mathcal{C}(d_A,d_W).
\end{equation}
Consequently, the normalized ratio is
\begin{align}
\frac{\psi_W(z_+)}
{\psi_W(z_+;\infty)}
&=
1-r_A r_W e^{-2\kappa_0L}.
\end{align}
The free energy of the scalar field is
\begin{equation}
     \mathcal F = \frac{k_BT}{2}  \int \frac{d^2k_\parallel}{(2\pi)^2} \sum_n \ln(1-r_A r_W e^{-2\kappa_0L}), 
\end{equation}
which is the standard Lifshitz expression.
A similar calculation can be done for the TM polarization contribution. 

\bibliographystyle{apsrev4-2}

\bibliography{ref}

\end{document}